\documentclass[a4paper,twoside,12pt]{article}
\usepackage{epsfig}
\newcommand{\eg}{{\em e.g.}}
\newcommand{\ie}{{\em i.e.}}
\newcommand{\cf}{{\em c.f.}}

\newcommand{\Pom}{I$\!$P}

\newcommand{\Pma}{I\!\!P}
\newcommand{\Reg}{I$\!$R}

\newcommand{\Rma}{I\!\!R}
\def\lsim{\mathrel{\rlap{\lower4pt\hbox{\hskip1pt$\sim$}}
    \raise1pt\hbox{$<$}}}         %less than or approx. symbol
\def\gsim{\mathrel{\rlap{\lower4pt\hbox{\hskip1pt$\sim$}}
    \raise1pt\hbox{$>$}}}         %greater than or approx. symbol
\begin{document}
\thispagestyle{empty}   % to suppress the page number on the first page
\noindent
%\hfill Draft  \today \\
DESY 99--009          \hfill ISSN 0418-9833\\
TSL/ISV-99-0202 \\
January 1999        %  \hfill {\bf DRAFT} \today\\
\vspace*{5mm}
\begin{center}
  \begin{Large}
  \begin{bf}
Diffractive Hard Scattering\footnote{Lectures at the
Advanced Study Institute on Techniques and Concepts of High Energy Physics,
\linebreak
St.\ Croix, USVI, 1998, to appear in the proceedings.}\\
  \end{bf}
  \end{Large}
  \vspace{5mm}
% \vspace{1cm}
  \begin{large}
    Gunnar Ingelman\footnote{ingelman@desy.de}\\  
  \end{large}
  \vspace{3mm}
Deutsches~Elektronen-Synchrotron~DESY,
Notkestrasse~85,~D-22603~Hamburg,~FRG\\
Dept. of Radiation Sciences, Uppsala University,
Box 535, S-751 21 Uppsala, Sweden\\
\vspace{5mm}
\end{center}
\begin{quotation}
\noindent
{\bf Abstract:}
Diffraction is an old subject which has received much interest in recent years 
due to the advent of diffractive hard scattering. We discuss some theoretical 
models and experimental results that have shown new striking effects, \eg \ 
rapidity gaps in jet and $W$ production and in deep inelastic scattering. 
Many aspects can be described through the exchange of a pomeron with a parton 
content, but the pomeron concept is nevertheless problematic. 
New ideas, \eg \ based on soft colour interactions, 
have been introduced to resolve these problems and provide a unified 
description of diffractive and non-diffractive events. This is part of the 
general unsolved problem of non-perturbative QCD and confinement. 
\end{quotation}
%

%%%%%%%%%%%%%%%%%%%%%%%%%%%%%%%%%%%%%%%%%%%%%%%%%%%%%%%%%%%%%%%%%%%%%%%%%%%%%%
%
\section{Introduction}\label{sec:Intro}
Ideas on diffraction have been developed over a long time. Quite old 
({\em `old-old'}) is the Regge approach \cite{Regge} with a pomeron mediating 
elastic and diffractive interactions \cite{diffraction}. 
Being from pre-QCD times, Regge phenomenology only considers soft interactions 
described in terms of hadrons. In a modern QCD-based language one would like 
to understand diffraction on the parton level. This was the starting point 
of the by now {\em `old new'} idea \cite{IS} that one 
should probe the structure of the pomeron through a hard scattering in 
diffractive events. By introducing a hard scale one should resolve partons 
in the pomeron and also make calculations possible through perturbative QCD
(pQCD). This opened the new branch of {\em diffractive hard scattering}
with models and the discovery by UA8 \cite{UA8} as discussed in section 
\ref{sec:idea}. 

The models are based on a factorization between the new concepts of a 
`pomeron flux' (in the proton) and a `pomeron structure function' in terms of 
parton density functions. These ideas may be interpreted as the pomeron being
analogous to a hadron (maybe a glueball?) as discussed in section 
\ref{sec:pomeron}. 

The  discovery of rapidity gap events in deep inelastic scattering (DIS) 
at HERA was a great surprise to most people, although it had been predicted
as a natural consequence of the diffractive hard scattering idea \cite{IS}. 
The pointlike probe in DIS makes it an ideal way to measure the parton 
structure of diffraction. This is discussed in section \ref{sec:HERA}
together with diffractive production of jets and $W$'s at the Tevatron. 
Although pomeron-based models may work phenomenologically, 
there are conceptual and theoretical problems as  
discussed in section \ref{sec:problems}. 

\begin{figure}[tbh]
\begin{center}
\epsfig{file=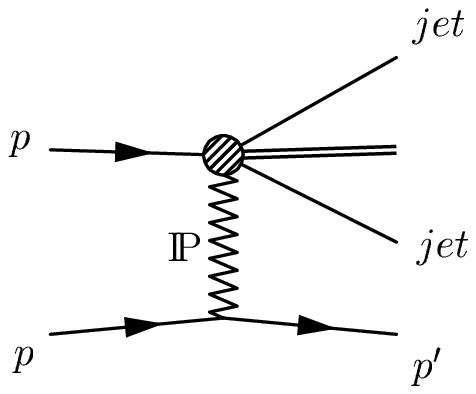,width=45mm} 
\epsfig{file=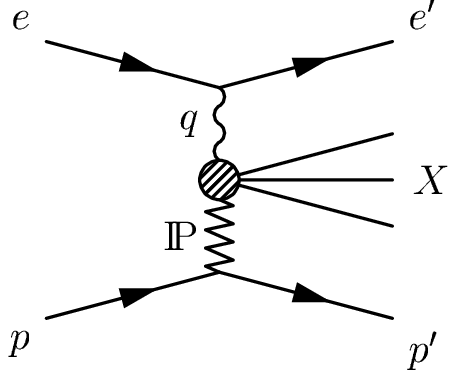,width=45mm} 
\epsfig{file=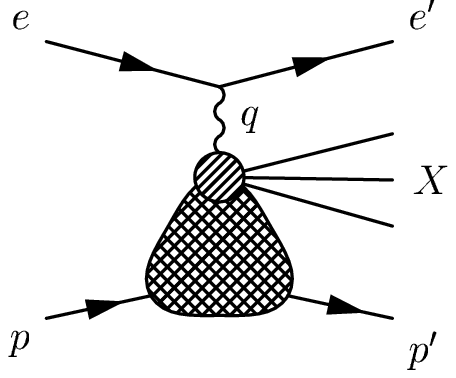,width=45mm} 
\end{center}
\caption{{\bf (a,b)} Diffractive hard scattering, in $p\bar{p}$ and DIS $ep$,
in the pomeron approach. {\bf (c)} Diffractive DIS illustrating the long
space-time scale for the soft interaction at the proton `vertex'.  
\label{fig:P-soft}}
\end{figure}

These problems are related to the general unsolved problem of non-perturbative
QCD (non-pQCD). Diffraction is one important aspect of this, others are 
hadronization in high energy collisions and the confinement of quarks and
gluons. In recent years there has been an increased interest for these 
problems and efforts are made based on new ideas and methods as discussed
in section \ref{sec:SCI}. The hard scale in diffractive hard scattering 
only solves part of the problem by making the upper part of the diagrams 
in Fig.~\ref{fig:P-soft} calculable in perturbation theory. 
However, the soft, lower part of the interaction occurs over a large 
space-time as illustrated in Fig.~\ref{fig:P-soft}c and must be 
treated with some novel non-pQCD methods. 

One such {\em `new-new'} idea is  
the {\em soft colour interaction} (SCI) model \cite{SCI}, which is an explicit 
attempt to describe non-pQCD interactions in a Monte Carlo event generation
model. Although it is quite simple, it is able to describe data on different
diffractive and non-diffractive interactions as discussed in section 
\ref{sec:SCI}. However, a better theoretical basis for this kind of 
models is certainly needed. In addition, the rapidity gaps between 
high-$p_\perp$ jets observed at the Tevatron are still a challenge to 
understand (section \ref{sec:jgj}). 
In conclusion (section \ref{sec:conclusions}),
although substantial progress has been made recently, diffractive scattering
is still a basically unsolved problem which provides challenges for the 
future.  
%

%%%%%%%%%%%%%%%%%%%%%%%%%%%%%%%%%%%%%%%%%%%%%%%%%%%%%%%%%%%%%%%%%%%%%%%%%%%%
% 
\section{Rapidity gaps and the pomeron concept}\label{sec:pomeron}
\begin{figure}[htb]
\begin{center}
\epsfig{file=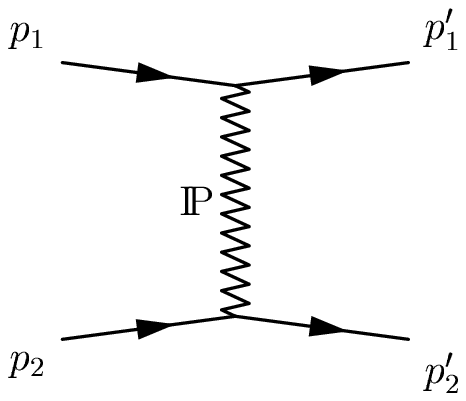,width=30mm}
\epsfig{file=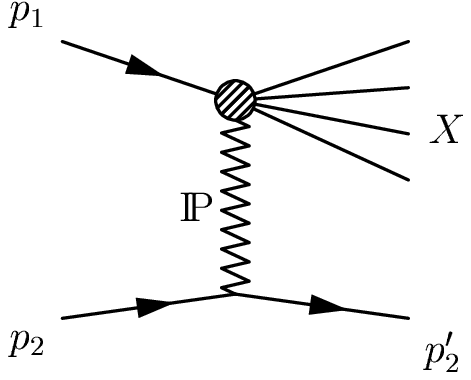,width=30mm}
\epsfig{file=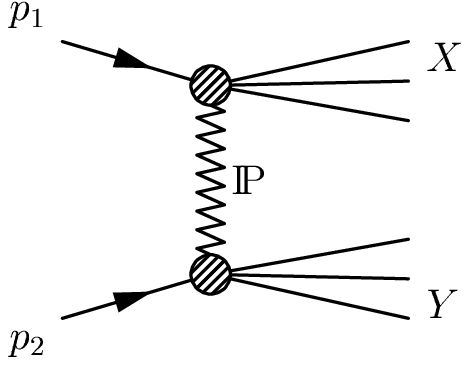,width=30mm}
\epsfig{file=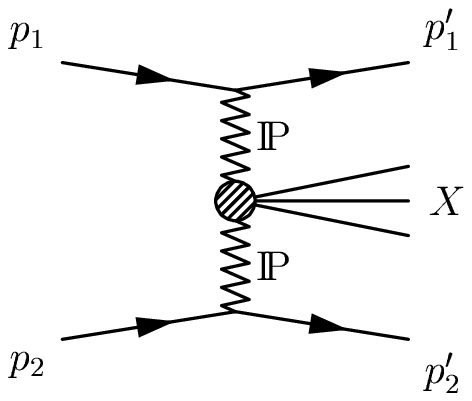,width=30mm}
\epsfig{file=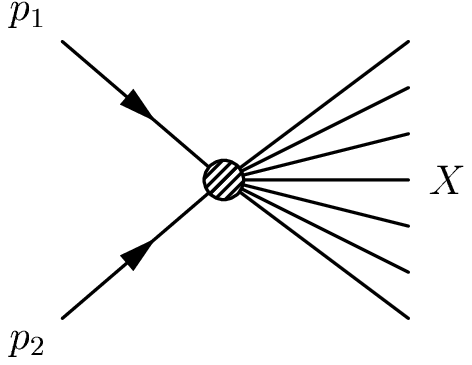,width=30mm}
\epsfig{file=diffractive.eps,width=150mm}
\end{center}
\caption{Rapidity distribution, $dN/dy$, of final state hadrons in 
hadron-hadron collisions with interpretations in terms of pomeron exchange:
(a) elastic, 
(b) single diffractive, (c) double diffractive, 
(d) double pomeron exchange and (e) totally inelastic interactions. 
\label{fig:rapgaps}}
\end{figure}

The dynamics of hadron-hadron interactions are largely not understood. 
Only the very small fraction of the cross section related to hard 
(large momentum transfer) interactions can be understood from first 
principles using pertubation theory, \eg \ jet production in QCD or
$\gamma^\star,W,Z$ production in electroweak theory. The large cross 
section (${\cal O}(mb)$) processes, on the other hand, are given by 
non-pQCD for which proper theory is lacking and only phenomenological 
models are available. These processes are classified in terms of their 
final states as illustrated in Fig.~\ref{fig:rapgaps}. 

The distribution of final state hadrons is then usually expressed in 
terms of the rapidity variables 
\begin{equation}\label{eq:rapidity}
{\rm rapidity}\;\;
y=\frac{1}{2}\ln{\frac{E+p_z}{E-p_z}} \approx -\ln{\tan{\frac{\theta}{2}}}
=\eta \;\; {\rm pseudorapidity} 
\end{equation}
where the approximation becomes exact for massless particles 
and the polar angle $\theta$ is with respect to the $z$-axis along the beam. 
In a totally inelastic interaction (Fig.~\ref{fig:rapgaps}e) 
the hadrons are distributed with a flat rapidity plateau. This 
corresponds to longitudinal phase space where the transverse momenta 
are limited to a few hundred MeV, but longitudinal momenta cover the 
available phase space. This is in accordance with hadronization models,
\eg \ the Lund string model \cite{Lund}, where 
longitudinal momenta are given by a scaling fragmentation function and
transverse momenta are strongly suppressed above the scale of soft 
interactions. The probability to have events with a gap, \ie \ a region 
without particles, due to statistical fluctuations in such a rapidity
distribution decreases exponentially with the size of the gap.  

Experimentally one observes a much higher rate of gaps. 
{\em Diffraction} is nowadays often defined as events with 
{\em large rapidity gaps which are not exponentially suppressed}
\cite{Bj}. 
This is, however, a wider definition than that previously often used 
in terms of a leading proton taking a large fraction (\eg \ $x_F\gsim 0.9$)
of the beam proton momentum which enforces a rapidity gap simply 
by kinematical constraints. However, a gap can be anywhere in the event 
and therefore allow a forward system of higher mass than a single proton. 
The definition chosen reflects what the experiments actually observe. 
The leading protons go down the beam pipe and their detection require 
tracking detectors in `Roman pots' which are moved into the beam 
pipe to cover the very small angles caused by the scattering itself or 
the bending out of the beam path by machine dipole magnets. 

The simplest gap events occur in elastic and single diffractive scattering
(Fig.~\ref{fig:rapgaps}a,b). Due to the scattered proton there is obviously 
an exchange of energy-momentum, but not of quantum numbers. 
In Regge phenomenology this is described as the exchange of an 
`object' with vacuum quantum numbers called a {\em pomeron} (\Pom ) after 
the Russian physicist Pomeranchuk. 
Regge theory \cite{Regge} is a description based on analyticity 
for scattering amplitudes in high energy interactions without large 
momentum transfers, but it is not a theory based 
on a fundamental Lagrangian like QCD.

The kinematics of single diffraction can be specified in terms of 
two variables, \eg \ the momentum fraction $x_p=p_f/p_i$ of the final proton
relative to the initial one and the momentum transfer $t=(p_i-p_f)^2$. 
The pomeron then takes the momentum fraction $x_{\Pma}=1-x_p$ and 
has a negative mass-squared $m_{\Pma}^2=t<0$ meaning that it is 
a virtual exchanged object. 
The other proton produces a hadronic system $X$ of mass $M_X^2=x_{\Pma}s$, 
\ie \ the invariant mass-squared of the `pomeron-proton collision'.
The cross section for single diffraction (SD) is experimentally found to be 
well described by
\begin{equation}\label{eq:dsigma_SD}
\frac{d\sigma_{SD}}{dt\, dx} \simeq \frac{1}{x_{\Pma}} 
% \left\{ a_1\exp{b_1t} + a_2\exp{b_2t} + \ldots \right\} 
  \left\{ a_1 e^{b_1t} + a_2 e^{b_2t} + \ldots \right\} 
  \simeq \frac{a}{M_X^2} \left| F(t) \right| ^2
\end{equation}
where the exponential damping in $t$ can be interpreted in terms of a proton 
form factor $F(t)$ giving the probability that the proton stays intact after 
the momentum `kick' $t$. 
With $x_{\Pma}<0.1$ the maximum $M_X$ reachable at ISR, S$p\bar{p}S$, Tevatron 
and LHC are 20, 170, 570 GeV and 4.4 TeV, respectively. However, the rate of
large $M_X$ events is suppressed due to the dominantly small pomeron momentum
fraction. This is the reason why it took until 1985 to demonstrate that the 
rapidity distribution of hadrons in the $X$-system shows longitudinal phase 
space \cite{longituinal}. 
Therefore, the pomeron-proton collision is similar to an ordinary 
hadron-proton interaction. This ruled out `fireball models' giving a 
spherically symmetric final state having a Gaussian rapidity distribution
\cite{diffraction}.   
Thus, the hadronic final state provides information on the interaction 
dynamics producing it.  

The Regge formalism relates the differential cross sections 
for different processes. 
This is achieved through the factorization of the different 
vertices such that the same kind of vertex in different processes is given 
by the same expression. The exchange of other than vacuum quantum numbers 
are described as, \eg , meson exchanges. Since the exchanged object is 
not a real state, but virtual with a negative mass-squared, it is actually 
a representation of a whole set of states (\eg \ mesons) with essentially
the same quantum numbers. The spin versus the mass-squared of such a set 
gives a linear relation which can be extrapolated to $m^2=t<0$ and provides
the trajectory $\alpha(t)$ for the exchange. This provides the essential 
energy dependence $\sigma \sim s^{2\alpha(t)-2}$ of the cross section. 
The pomeron 
trajectory $\alpha_{\Pma}(t)=1+\epsilon+\alpha 't \simeq 1.08+0.25t$ has 
the largest value of all trajectories 
at $t=0$ (intercept) which leads to the dominant 
contribution to the hadron-hadron cross section. 
Contrary to the $\pi$ and $\rho$ 
trajectories, which have well known integer spin states at the pole 
positions $t=m^2_{meson}$, there are no real states on the pomeron 
trajectory. However, a recently found spin-2 glueball candidate with mass
$1926\pm 12$ MeV \cite{WA91} fits well on the pomeron trajectory. 
% \cite{Landshoff_glueball}. 

This would be in accord with the suggestion that the pomeron is some 
gluonic system \cite{Low-Nussinov} which may be interpreted as a virtual 
glueball \cite{Simonov}. 
In a modern QCD-based language it is natural to consider a 
{\em pomeron-hadron analogy} where the pomeron is a hadron-like 
object with a quark and gluon content. Pomeron-hadron 
interactions would then resemble hadron-hadron collisions and give 
final state hadrons in longitudinal phase space, just as observed. 
There was, however, another view in terms of a {\em pomeron-photon analogy} 
\cite{pomeron-photon} where the pomeron is considered to have an effective pointlike 
coupling to quarks. Single diffractive scattering would then be similar to 
deep inelastic 
scattering and the exchanged pomeron scatters a quark out of the proton, 
leading to a longitudinal phase space after hadronization. This fits 
well with the experimental evidence for pomeron single-quark interactions
\cite{singelquark}. 

%%%%%%%%%%%%%%%%%%%%%%%%%%%%%%%%%%%%%%%%%%%%%%%%%%%%%%%%%%%%%%%%%%%%%%%%%%%%%%
%
\section{Idea and discovery of diffractive hard scattering}\label{sec:idea}
To explore the diffractive interaction further, we \cite{IS} 
introduced in 1984 the new idea that one should use a hard scattering 
process to probe the pomeron interaction at the parton level. 
In retrospect this seems obvious and simple, but at that time it was 
quite radical and was criticised. The idea was launched before the 
observations of longitudinal event structure in diffraction, the glueball
candidate on the pomeron trajectory and the pomeron single-quark interactions
discussed above. 
Furthermore, diffraction was at that time a side issue in particle 
physics that was ignored by most people. 
\begin{figure}[htb]
\begin{center}
\epsfig{file=feyn_pp_pom.eps,width=50mm} 
\epsfig{file=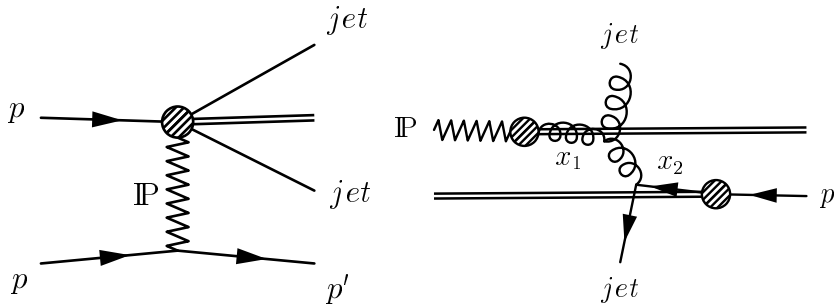,width=60mm} 
\end{center}
\caption{Single diffractive scattering with pomeron exchange giving a 
pomeron-proton interaction with a hard parton level subprocess 
producing jets in the $X$-system. 
\label{fig:IS}}
\end{figure}

Based on the pomeron factorization hypothesis, the diffractive hard 
scattering process was considered \cite{IS} in terms of an exchanged pomeron 
and a pomeron-particle interaction were a hard scattering process on
the parton level may take place as illustrated in Fig.~\ref{fig:IS}.
The diffractive hard scattering 
cross section can then be expressed as the product of the inclusive 
single diffractive cross section and the ratio of the pomeron-proton 
cross sections for producing jets and anything, \ie 
\begin{equation}\label{eq:IS}
\frac{d\sigma_{jj}}{dtdM_X^2} = \frac{d\sigma_{SD}}{dtdM_X^2}\; 
\frac{\sigma(\Pma p\to jj)}{\sigma(\Pma p\to X)}
\end{equation}
Here, $d\sigma_{SD}$ can be taken as the parametrization of data in 
eq.~(\ref{eq:dsigma_SD}) and the total pomeron-proton cross section
$\sigma(\Pma p\to X)$ can be extracted from data using the Regge formalism
resulting in a value of order 1 mb. Together these parts of eq.~(\ref{eq:IS}) 
can be seen as an expression for a {\em pomeron flux} 
$f_{\Pma /p}(x_{\Pma},t)$ in the beam proton. 
The cross section for pomeron-proton to jets, $\sigma(\Pma p\to jj)$,
is assumed to be given by pQCD as 
\begin{equation}\label{eq:IS-QCD}
\sigma(\Pma p\to jj) = \int dx_1\, dx_2\, d\hat{t} \sum_{ij} \;
f_{i/\Pma}(x_1,Q^2) f_{j/p}(x_2,Q^2) \frac{d\hat{\sigma}}{d\hat{t}}
\end{equation}
where a parton density function $f_{i/\Pma}$ for the pomeron is introduced
in analogy with those for ordinary hadrons. The pomeron parton density 
functions were basically unknown, but assuming the pomeron to be gluon 
dominated it was resonable to try $xg(x)=ax(1-x)$ or $xg(x)=b(1-x)^5$
for the cases of only two gluons or of many gluons similar to the proton.
Similarly, if the pomeron were essentially a $q\bar{q}$ system one would 
guess $xq(x)=cx(1-x)$. The normalisation constants $a,b,c$ can be chosen 
to saturate the momentum sum rule $\int_0^1 dx \sum_i xf_{i/\Pma}(x)=1$, 
which seems like a reasonable assumption to get started. 

This formalism allows numerical estimates for diffractive hard scattering
cross sections. Diffractive jet cross sections at the CERN S$p\bar{p}$S
collider energy were found \cite{IS} to be large enough to be observable. 
Furthermore, turning the formalism into a Monte Carlo (MC)
program (precursor to {\sc Pompyt} \cite{Pompyt} described below)
to simulate complete 
events, demonstrated a clearly observable event signature: 
a leading proton ($x_F\gsim 0.9$) separated by a large rapidity gap 
from a central hadronic system with high-$p_{\perp}$ jets. 

Based on these predictions, the UA8 experiment was approved and constructed. 
It had Roman pots in the beam pipes to measure the momentum of leading 
(anti)protons and used the UA2 central detector to observe jets. 
The striking event signature were observed in 1987 \cite{UA8}
signalling the discovery of the diffractive hard scattering phenomenon, 
which was investigated further with more data 
\cite{UA8_superhard,UA8_cross-section}.

\begin{figure}[b]
\begin{center}
\epsfig{file=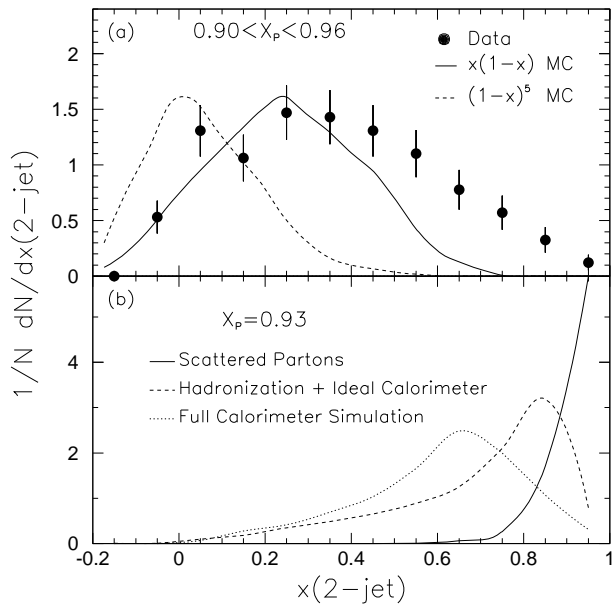,width=80mm, 
   bbllx=25pt,bblly=35pt,bburx=505pt,bbury=515pt,clip=}
\end{center}
\vspace*{-7mm}
\caption{%
(a) Distribution of scaled longitudinal momentum $x_F$ for the two-jet
system in diffractive events defined by a leading proton ($0.90<x_p<0.96$). 
Data from UA8 compared to the Monte Carlo model using 
the indicated hard and soft parton densities in the pomeron. 
(b) Monte Carlo results at the parton level, after hadronization and after 
detector simulation assuming `super-hard' partons in the pomeron
($xf(x)\sim \delta(1-x)$). From \protect\cite{UA8_superhard}.}  
\label{fig:superhard}
\end{figure}

The observed jets showed the characteristic properties of QCD jets 
as quantified in the Monte Carlo, \eg \ jet $E_\perp$ 
and angular distributions and energy profiles. 
The longitudinal momentum of the jets gives information on the momentum 
fraction ($x_1$ in Fig.~\ref{fig:IS}b) of the parton in the pomeron;  
a change in the shape of the $x_1$-distribution shifts the parton-parton cms 
with respect 
to the $X$ cms and thereby the momentum distribution of the jets \cite{IS}. 
Comparison of data and the Monte Carlo shows a clear preference for 
a hard parton distribution \cite{UA8_superhard}. Using a quark or gluon 
distribution $xf(x)\sim x(1-x)$ gives a resonable description of the 
observed $x_F$-distribution of the jets, although giving too little in the 
tail at large $x_F$. This is more clearly seen, if 
instead of considering individual jets, one takes both jets in each event 
and plot the longitudinal momentum of this pair, Fig.~\ref{fig:superhard}. 
The excess at large $x_F$ can be described by having 30\% of the pomeron 
structure function in terms of a {\em super-hard} component with partons 
taking the entire pomeron momentum, \ie \ $xf(x)\sim \delta(1-x)$. 
The $\delta$-function can be seen as a representation of some more physical 
distribution which is very hard, \eg \ $xf(x)\sim 1/(1-x)$. 

With the UA8 data alone, one cannot distinguish between gluons or quarks
in the pomeron. The UA1 experiment has given some evidence for diffractive 
bottom production \cite{UA1}. This may be interpreted with a gluon-dominated 
pomeron such that the $gg\to b\bar{b}$ subprocess can be at work, but no firm 
conclusion can be made given the normalization uncertainty in the model and 
the experimental errors \cite{Bruni_Marseille}. 

UA8 have recently provided the absolute cross section for diffractive jet 
production \cite{UA8_cross-section}. 
This shows that, although the Monte Carlo model reproduces the shapes 
of various distributions, it overestimates the absolute cross section;
$\sigma(data)/\sigma(model)=0.30\pm 0.10$ or $0.56\pm 0.19$ for the 
model with the pomeron as a gluonic or a $q\bar{q}$ state, respectively. 
This have raised questions concerning the normalization of the pomeron 
flux and the pomeron structure function, as will be discussed in 
section \ref{sec:problems}. 

In summary, diffractive hard scattering has been discovered by UA8 and 
the main features can be interpreted in terms of an exchanged pomeron with 
a parton structure. 

%%%%%%%%%%%%%%%%%%%%%%%%%%%%%%%%%%%%%%%%%%%%%%%%%%%%%%%%%%%%%%%%%%%%%%%%%%%%%
% 
\section{Rapidity gap events at HERA and the Tevatron}\label{sec:HERA}
The above model for diffractive hard scattering can be naturally extended
to other kinds of particle collisions, $a+p\to p+X$ where $a$ can not only 
be any hadron but also a lepton or a photon.  
Based on the pomeron factorization hypothesis \cite{IS,BCSS} the cross section 
is $d\sigma (a+p\to p+X) = f_{\Pma/p}(x_{\Pma},t) \, d\sigma (a+\Pma \to X)$. 
The pomeron flux can be taken as a simple parametrization of data in terms of 
exponentials as above, or obtained from Regge phenomenology in the form 
\cite{DL}
\begin{equation}\label{eq:pomeron-flux}
f_{\Pma /p}(x_{\Pma},t)=\frac{9\beta_0^2}{4\pi^2}
\left( \frac{1}{x_{\Pma}}\right) ^{2\alpha_{\Pma}(t)-1} \left[ F_1(t)\right]^2
\end{equation}
with parameters obtained from data on hadronic inclusive diffractive scattering.  
Here, $\beta =3.24\: GeV^2$ is the mentioned effective pomeron-quark coupling
and $F_1(t) = (4m_p^2 -At)/(4m_p^2-t)\cdot (1-t/B)^{-2}$ is a proton 
form factor with $m_p$ the proton mass and parameters $A=2.8,\: B=0.7$. 
The pomeron trajectory is $\alpha_{\Pma}(t)\simeq 1.08+0.25t$. 

For the hard scattering cross section $d\sigma (a+\Pma \to X)$ one should 
use the relevant convolution of parton densities and parton cross sections,
\eg \ eq.~(\ref{eq:IS-QCD}) for hadron-pomeron collisions. 
In order to simulate complete events this formalism has been included 
in the Monte Carlo program {\sc Pompyt} \cite{Pompyt} based on the 
Lund Monte Carlo {\sc Pythia} \cite{Pythia}. 
In particular, there are options for different 
pomeron flux factors and parton densities. Moreover, {\sc Pompyt} also
contains pion exchange processes where a pion, with a flux factor and 
parton densities, replaces the pomeron as an example of other possible 
Regge exchanges.  

\subsection{Diffractive DIS at HERA}\label{subsec:HERA}

As suggested already in \cite{IS}, one should 
probe the pomeron structure with deep inelastic scattering, \eg \ at HERA.
The advantage would be to have a clean process with a well understood 
point-like probe with high resolving power $Q^2$. The experimental 
signature should be clear; a quasi-elastically scattered proton (going down 
the beam pipe) well separated by a rapidity gap from the remaining hadronic
system. The kinematics is then described by the diffractive variables 
$x_{\Pma}$ (or $x_p=1-x_{\Pma}$) and $t$, as above, and the standard DIS 
variables $Q^2=-q^2=-(p_e-p_{e'})^2$ and Bjorken $x=Q^2/2P\cdot q$ 
(where $P,p_e,p_{e'},q$ are the four-momenta of the initial proton, 
initial electron, scattered electron and exchanged photon, respectively).  

The cross section for diffractive DIS can then be written \cite{Ingelman-Prytz}
\begin{equation}\label{eq:sigmaDIS}
\frac{d\sigma (ep\to epX)}{dxdQ^2dx_{\Pma}dt} = \frac{4\pi \alpha^2}{xQ^4}\, 
\left( 1-y+\frac{y^2}{2} \right) \, F_2^D (x,Q^2;x_{\Pma} ,t)
\end{equation}
where the normal proton structure function $F_2$ has been replaced by a 
corresponding diffractive one, $F_2^D$, with $x_{\Pma}$ and $t$ specifying the 
diffractive conditions. 
Only the dominating electromagnetic interaction 
is here considered and $R=\sigma_L/\sigma_T$ is neglected for simplicity. 
If pomeron factorization holds, then $F_2^D$ can be factorized into a 
pomeron flux and a pomeron structure function, \ie \
$F_2^D (x,Q^2;x_{\Pma} ,t) = f_{\Pma/p}(x_{\Pma},t)\, F_2^{\Pma}(z,Q^2)$ where 
the pomeron structure function 
$F_2^{\Pma}(z,Q^2)=\sum_f e_f^2\left( zq_f(z,Q^2)+z\bar{q}_f(z,Q^2)\right)$
is given by the densities of (anti)quarks of flavour $f$ and with a fraction 
$z=x/x_{\Pma}$ of the pomeron momentum. 
Since the photon does not couple directly to gluons, they will only enter indirectly 
through $g\to q\bar{q}$ as described by QCD evolution or the photon-gluon 
fusion process.

\begin{figure}[ht]
\begin{center}
\epsfig{file=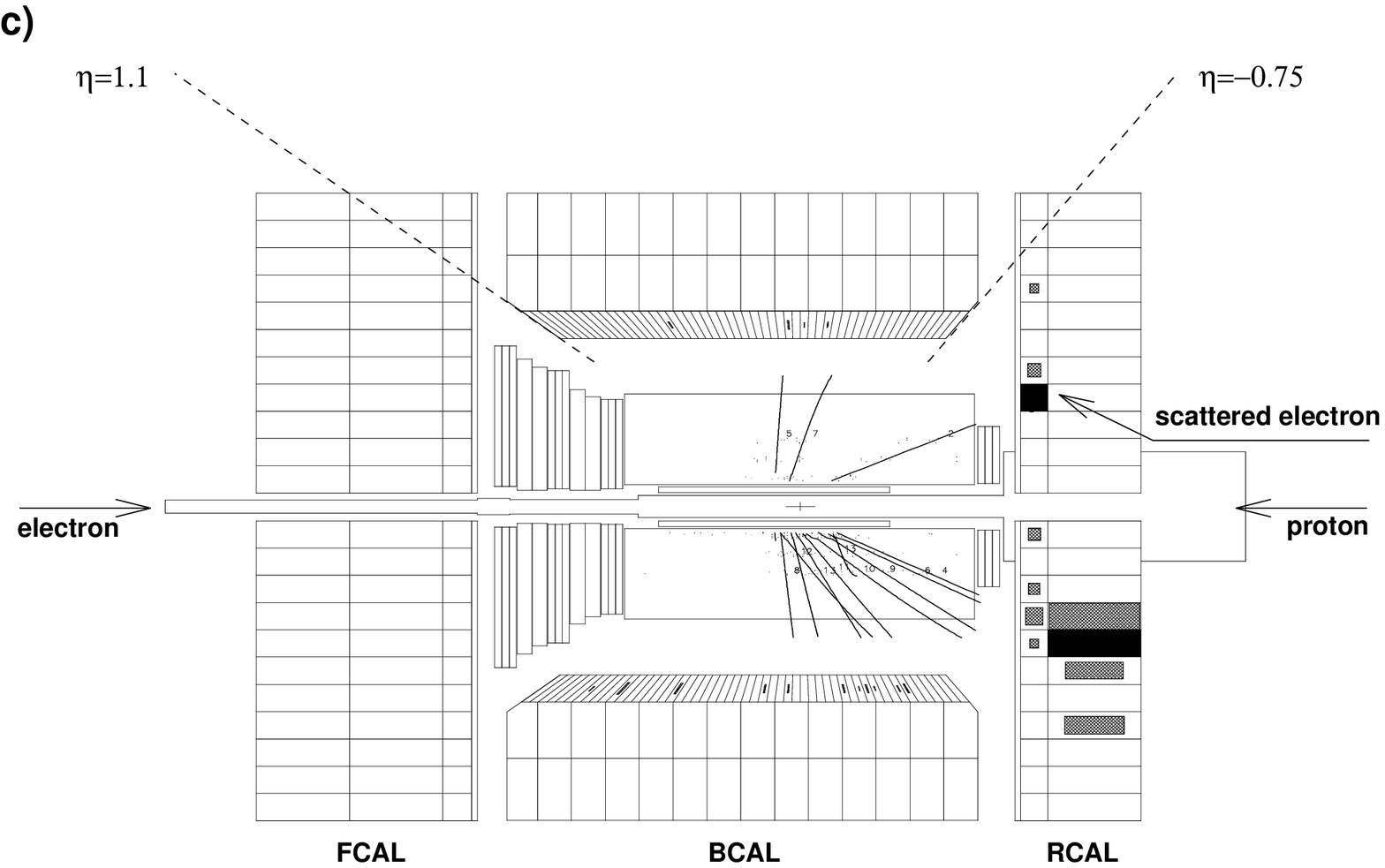,width=80mm,clip=}
%   bbllx=6pt,bblly=0pt,bburx=570pt,bbury=348pt,clip=}
\epsfig{file=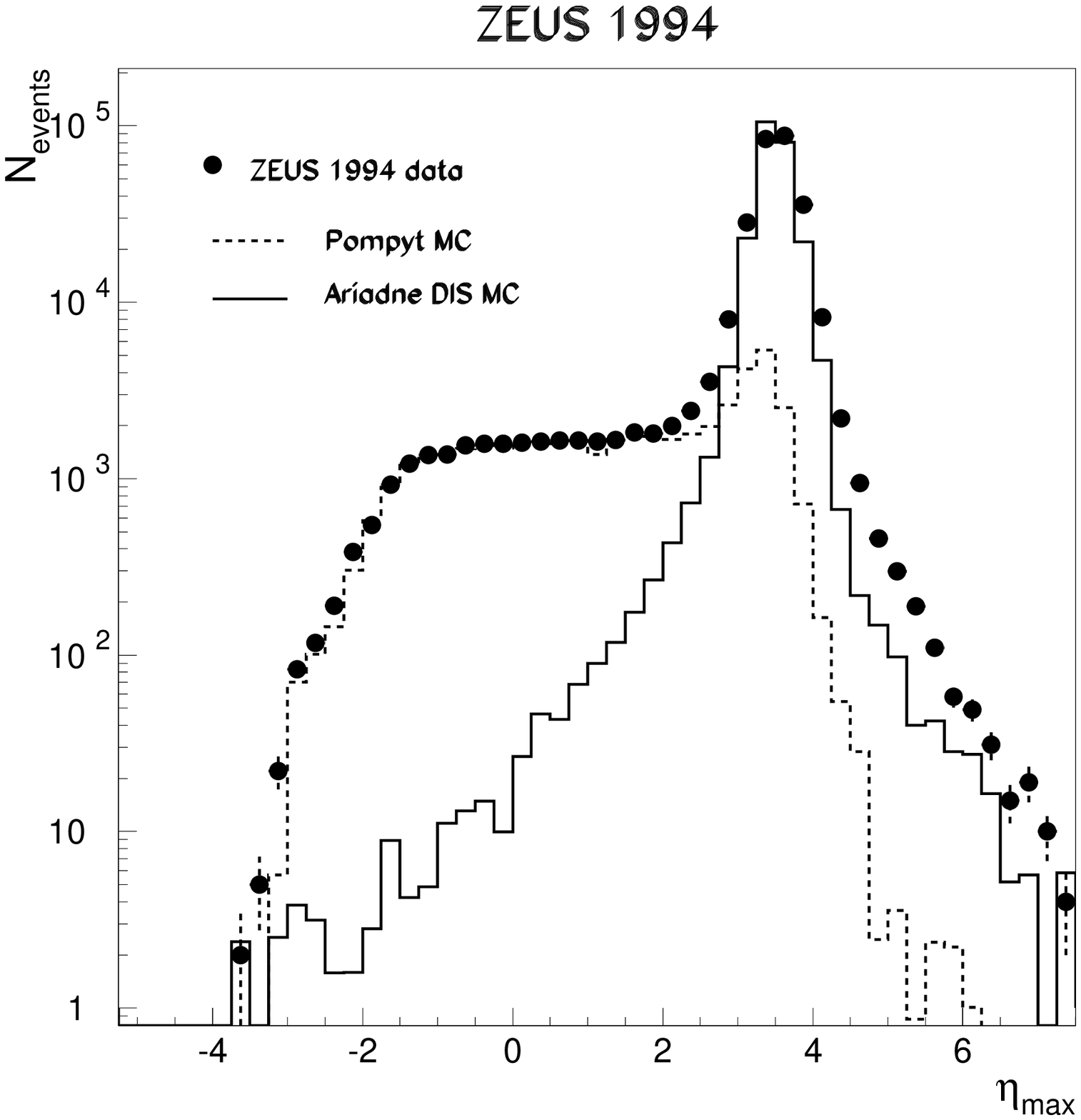,width=70mm, 
   bbllx=20pt,bblly=155pt,bburx=525pt,bbury=645pt,clip=}
\end{center}
\caption[junk]{{\bf (a)} The ZEUS detector with a rapidity gap event, \ie \ 
having no tracks or energy depositions in the forward (proton beam direction) 
part of the detector. 
\linebreak
{\bf (b)} Distribution of $\eta_{max}$, the maximum pseudorapidity of observed 
tracks/energy, in ZEUS data compared to Monte Carlo simulations using 
{\sc Ariadne} \cite{Ariadne} (full histogram) for ordinary DIS 
and {\sc Pompyt} \cite{Pompyt} (dashed) for DIS on a pomeron with a 
hard quark density ($zq(z)\sim z(1-z)$), 
with normalisation adjusted such that the sum fits the data. 
Courtesy of the ZEUS collaboration. 
\label{fig:gap_ZEUS_etamax}}
\end{figure}

Although diffractive DIS had been predicted in this way 
\cite{IS,Ingelman-Prytz,DDISpred}, 
it was a big surprise to many when it was observed first by ZEUS 
\cite{ZEUS_gaps} and then by H1 \cite{H1_gaps}. 
Since leading proton detectors were not available at that time, 
it was the large rapidity gap that was the characteristic observable, 
\ie \ no particle or energy depositions in the forward part of the detector
as shown in Fig.~\ref{fig:gap_ZEUS_etamax}a.
Leading protons have later been clearly observed 
\cite{leading-pn}, 
but the efficiency is low so the dominant diffractive data 
samples are still defined in terms of rapidity gaps. 
A simple observable to characterize the effect is $\eta_{max}$ giving, 
in each event, the maximum pseudo-rapidity where an energy deposition is 
observed. 
Fig.~\ref{fig:gap_ZEUS_etamax}b shows the distribution of this quantity.

Although the bulk of the data with $\eta_{max}$ in the forward region is well
described by ordinary DIS Monte Carlo events, there is a large excess with 
a smaller $\eta_{max}$ corresponding to the central region or even in the 
electron hemisphere. This excess is well described by {\sc Pompyt} as 
deep inelastic scattering on an exchanged pomeron with a hard quark density.
The gap events have the same $Q^2$ dependence as normal 
DIS and are therefore not some higher twist correction. Their overall rate is
about 10\% of all events, so it is not a rare phenomenon. 
\begin{figure}[htb]
\begin{center}
\epsfig{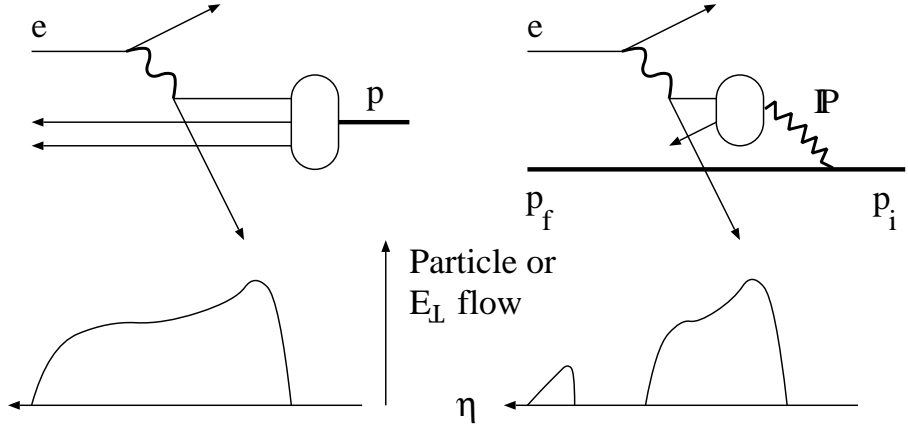}
%   bbllx=55pt,bblly=510pt,bburx=415pt,bbury=700pt,clip=}
\end{center}
\caption[junk]{Illustration of (a) normal DIS and (b) diffractive DIS 
at HERA and the 
resulting particle or transverse energy flow in laboratory pseudorapidity.
\label{fig:gap_graph}}
\end{figure}

In normal DIS, a quark is scattered from the proton leaving a colour charged 
remnant (diquark in the simplest case). This gives rise to a colour field 
(\eg \ a string) between the separated colour charges, such that the 
hadronization 
gives particles in the whole intermediate phase space region as 
illustrated in Fig.~\ref{fig:gap_graph}a. 
The gap events correspond to the scattering on a colour singlet object, 
Fig.~\ref{fig:gap_graph}b, which gives no colour field between the hard 
scattering system and the proton remnant system. 
Therefore, no hadrons are produced in the region 
between them, \ie \ a rapidity gap appears.
The size of the gap is basically a kinematic effect. The larger fraction
of the proton beam momentum that is carried by the forward going colour 
singlet proton remnant system, the smaller fraction remains for other particles 
which therefore emerge at smaller rapidity. 
The forward going system $Y$ must have a small invariant mass in order to 
escape undetected in the beam pipe. It is mostly a proton with a large fraction
of the beam momentum and only a very small angular deflection. 

\begin{figure}[p]
\begin{center}
\epsfig{file=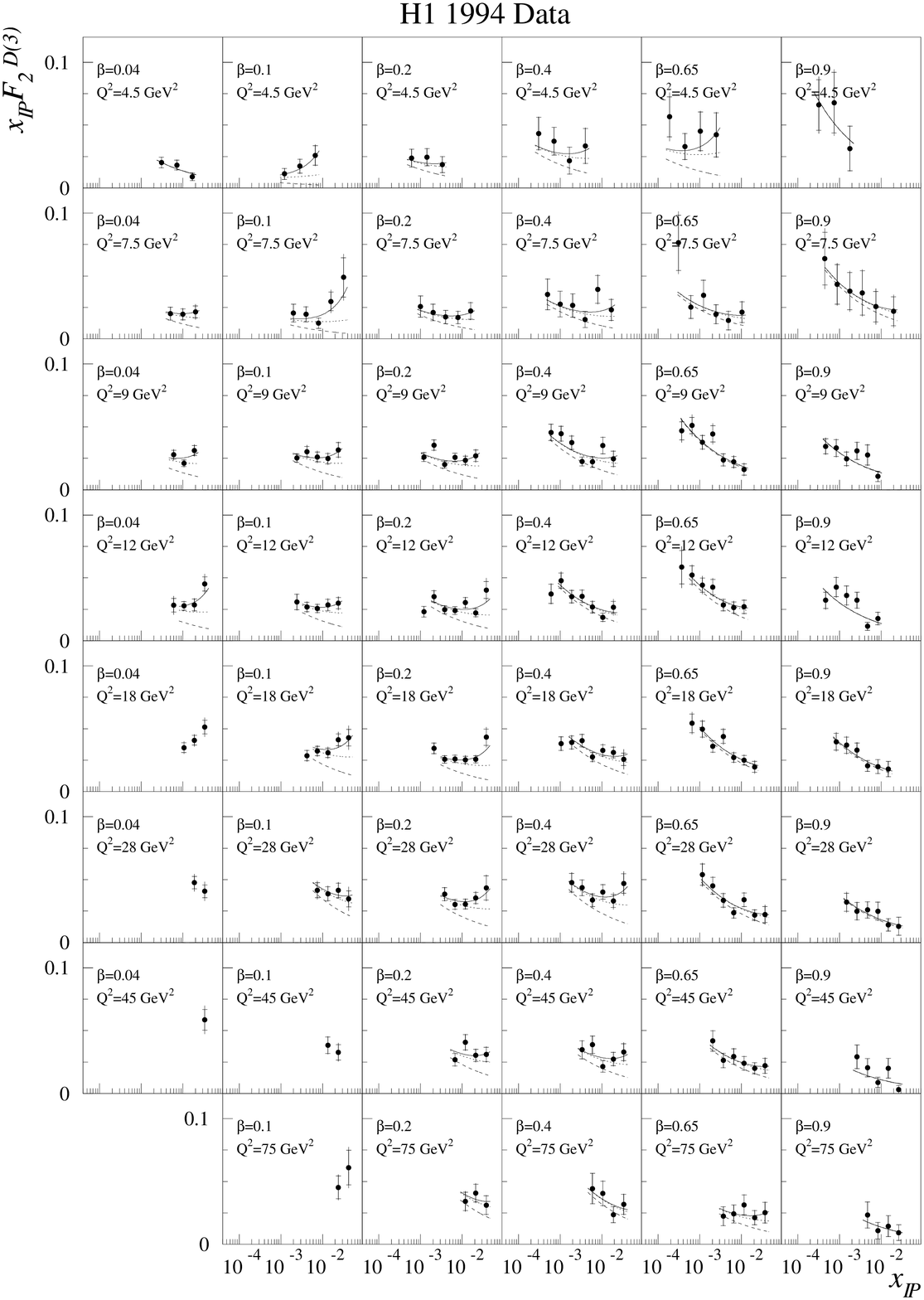,width=150mm, 
   bbllx=15pt,bblly=10pt,bburx=800pt,bbury=1110pt,clip=}
\end{center}
\caption[junk]{The diffractive structure function 
$x_{\Pma} F_2^{D(3)}$ for bins in $\beta$ and $Q^2$.
H1 data with Regge fits for pomeron and one reggeon exchange 
with their interference (full curves), only pomeron exchange (dashed curves)
and pomeron exchange plus the interference (dotted curves) \cite{H1_newF2D}.
\label{fig:F2D3}}
\end{figure}

Since the $Y$ system is not observed the $t$ variable is not measured, 
but is usually negligibly small (\cf \ the proton form factor above). 
However, with the invariant definitions 
\begin{eqnarray}\label{eq:invariantdef}
x_{\Pma} & = & \frac{q\cdot (P-p_Y)}{q\cdot P} = 
 \frac{Q^2+M_X^2-t}{Q^2+W^2-m_p^2} \approx \frac{x(Q^2+M_X^2)}{Q^2} \\
z=\beta & = & x/x_{\Pma}=\frac{-q^2}{2q\cdot(P-p_Y)}=\frac{Q^2}{Q^2+M_X^2-t}
\approx \frac{Q^2}{Q^2+M_X^2} 
\end{eqnarray}
$x_{\Pma}$ can be reconstructed from the DIS variables and the 
$X$-system. Likewise, $z$ (or $\beta$) 
can be measured and corresponds to Bjorken-$x$ for 
DIS on the pomeron and can therefore be interpreted as 
the momentum fraction of the parton in the pomeron.

From the measured cross section of rapidity gap events, the diffractive
structure function $F_2^D$ can be extracted based on eq.~(\ref{eq:sigmaDIS}). 
Since $t$ is not measured it is effectively integrated out giving the 
observable $F_2^{D(3)}(x_{\Pma},\beta,Q^2)$. 
To a first approximation it was found 
\cite{H1_gaps} that the $x_{\Pma}$ dependence factorises and is of the 
form $1/x_{\Pma}^{n}$ with $n=1.19\pm 0.06 \pm 0.07$. This is in basic 
agreement with the expectation 
$f_{\Pma/p}\sim 1/x_{\Pma}^{2\alpha_{\Pma}(t)-1}\simeq 1/x_{\Pma}^{1.16+0.5t}$
from the pomeron Regge trajectory above. 

However, with the increased statistics and kinematic range available in 
the new data \cite{H1_newF2D} displayed in Fig.~\ref{fig:F2D3}, 
deviations from such a universal factorisation are observed. 
The power of the $x_{\Pma}$-dependence is found to depend on $\beta$. 
One way to interpret this is to introduce a subleading 
reggeon (\Reg ) exchange with expected trajectory 
$\alpha_{\Rma}(t)\simeq 0.55+0.9t$ and quantum numbers of the 
$\rho , \omega , a$ or $f$ meson \cite{H1_newF2D}. 
Fits to the data (Fig.~\ref{fig:F2D3})
show that although the pomeron still dominates, the meson exchange contribution
is important at larger $x_{\Pma}$ and causes $F_2^D$ to decrease slower 
(or $x_{\Pma} F_2^{D(3)}$ to even increase). The fit gives the intercepts 
$\alpha_{\Rma}(0)=0.50\pm 0.19$, in agreement with the expectation,
and $\alpha_{\Pma}(0)=1.20\pm 0.04$ which is, however, significantly larger 
than 1.08 obtained from soft hadronic cross sections. 
\begin{figure}[hp]
\begin{center}
\epsfig{file=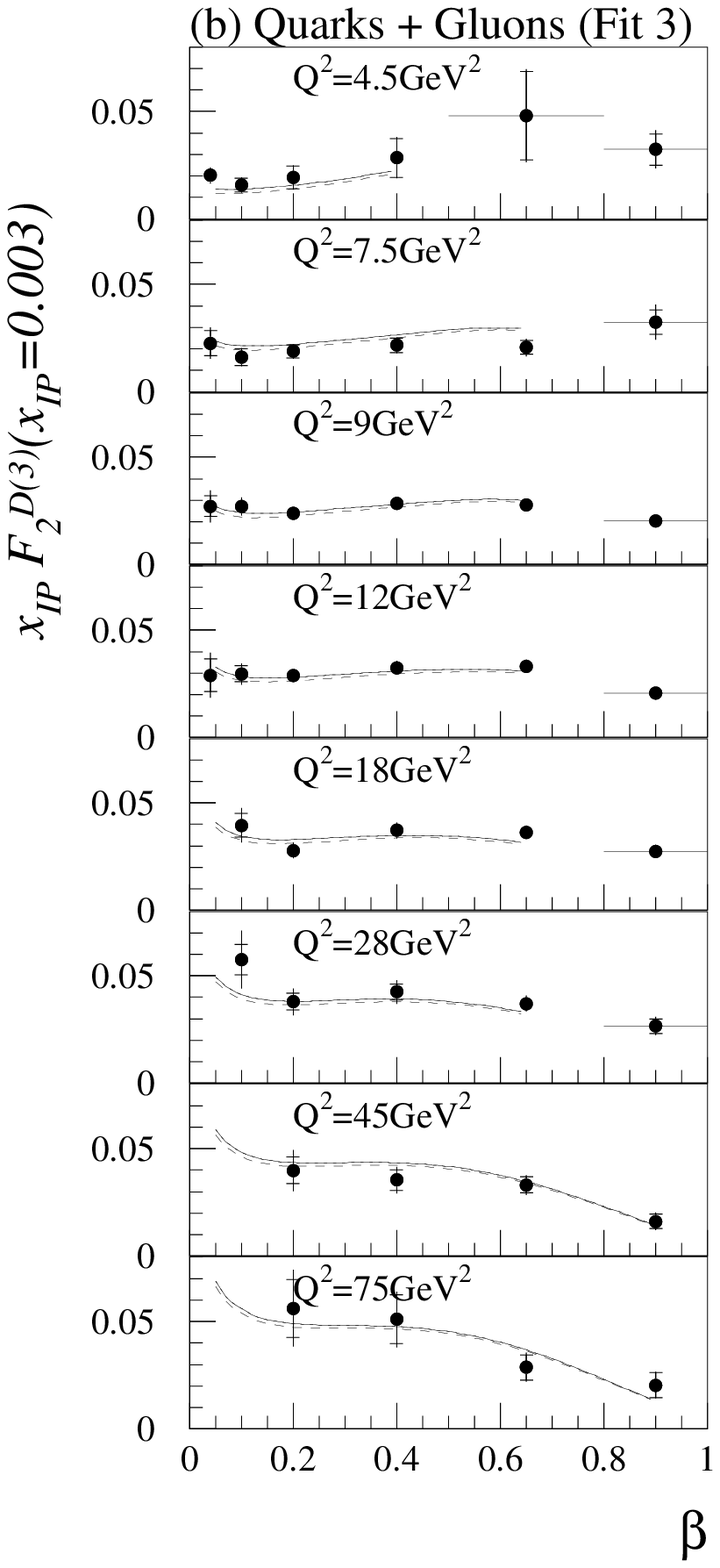,height=160mm}
\hspace*{-155mm}
\epsfig{file=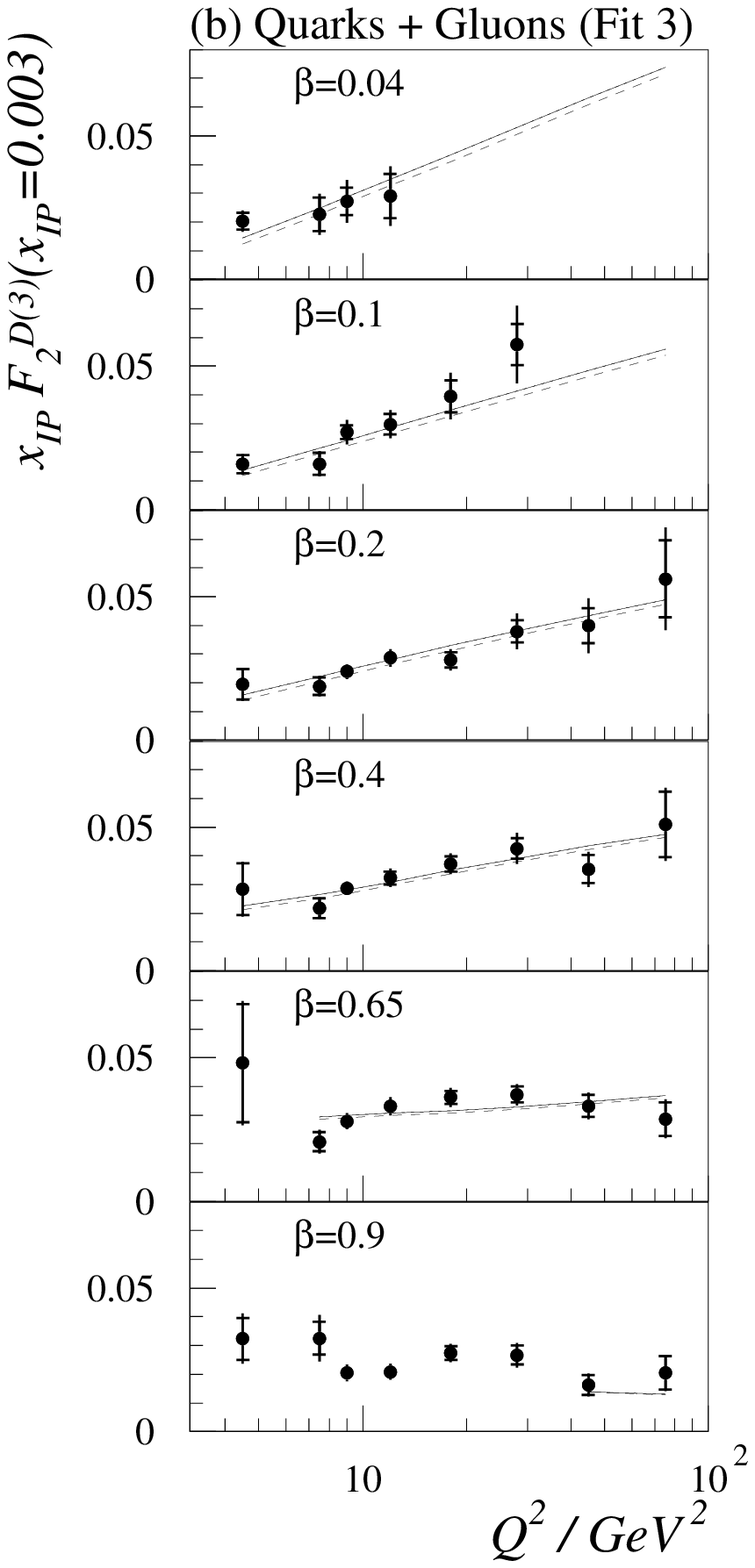,height=160mm}
\end{center}
\caption[junk]{H1 data on $F_2^D(z=\beta,Q^2)$ (interpolated to 
 $x_{\Pma}=0.003$), \ie \ the structure function of the exchanged 
 colourless object. The curves are fits of quark and gluon densities 
 with DGLAP evolution in QCD for pomeron exchange only (dotted) and 
 including the Reggeon exchange (full) \cite{H1_newF2D}.
\label{fig:F2Dtildeqg}}
\end{figure}

There is no evidence for a $\beta$ or $Q^2$ dependence in these intercepts 
and one can therefore 
integrate over $x_{\Pma}$ (using data and the fitted 
dependence), resulting in the measurement of $F_2^D(\beta,Q^2)$ 
shown in Fig.~\ref{fig:F2Dtildeqg}. 
Following the above framework, this quantity can be interpreted as the 
structure function of the exchanged colour singlet object, which is 
mainly the pomeron. The fact that $F_2^D$ is essentially scale
independent, \ie \ almost constant with $Q^2$, shows that the scattering occurs 
on point charges. The small $Q^2$ dependence present is actually compatible 
with being logarithmic as in normal QCD evolution, although the rise with 
$lnQ^2$ persists up to large values of $\beta$ in contrast to the proton 
structure function. There is only a weak dependence on $\beta$ such that 
the partons are quite hard and there is no strong decrease at 
large momentum fraction which is characteristic for ordinary hadrons. 

\begin{figure}[htb]
\begin{center}
\epsfig{file=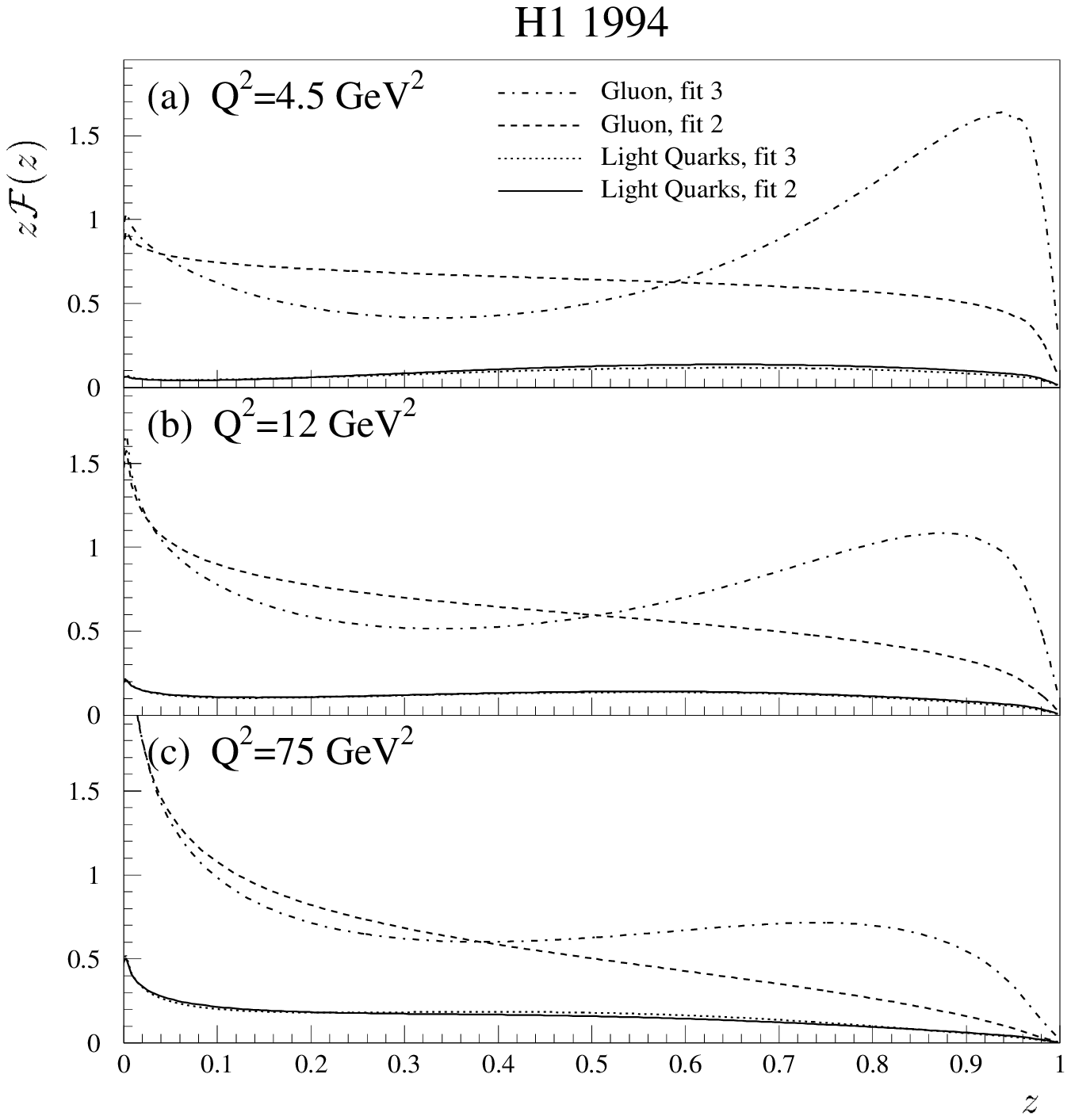,width=100mm}
\end{center}
\vspace*{-8mm}
\caption[junk]{Momentum-weighted distributions in fractional momenta 
$z$ of gluons and light quarks in the exchanged colorless object;
obtained at different $Q^2$ from a NLO DGLAP fit to 
$F_2^D(\beta,Q^2)$ in Fig.~\ref{fig:F2Dtildeqg}. 
From \cite{H1_newF2D}.
\label{fig:pom_partons}}
\end{figure}

These features are in accordance with 
a substantial gluon component in the structure of the diffractive exchange, 
as confirmed by a quantitative QCD analysis \cite{H1_newF2D}. 
Standard next-to-leading order DGLAP evolution \cite{DGLAP}
gives a good fit of 
$F_2^D(\beta,Q^2)$ as demonstrated in Fig.~\ref{fig:F2Dtildeqg}.
The fitted momentum distributions of quarks and gluons in the pomeron are 
shown in Fig.~\ref{fig:pom_partons}. Clearly, the gluon dominates and carries 
80--90\%  of the pomeron momentum depending on $Q^2$.
At low $Q^2$ the gluon distribution may even be peaked at large momentum 
fractions, \cf \ the superhard component observed by UA8 
\cite{UA8_superhard}, but when evolved to larger $Q^2$ it then becomes flatter 
in $\beta$. 

The general conclusion from these HERA data is therefore that the concept 
of an exchanged pomeron with a parton density seems appropriate. 
Moreover, Monte Carlo models, like {\sc Pompyt} \cite{Pompyt} and {\sc Rapgap} 
\cite{Rapgap} (which is also based on the above pomeron formalism), 
can give a good description of the observed rapidity gap events. 

\newpage
\subsection{Diffractive $W$ and jets at the Tevatron}\label{subsec:Tevatron}
\begin{figure}[hbt]
\begin{center}
\epsfig{file=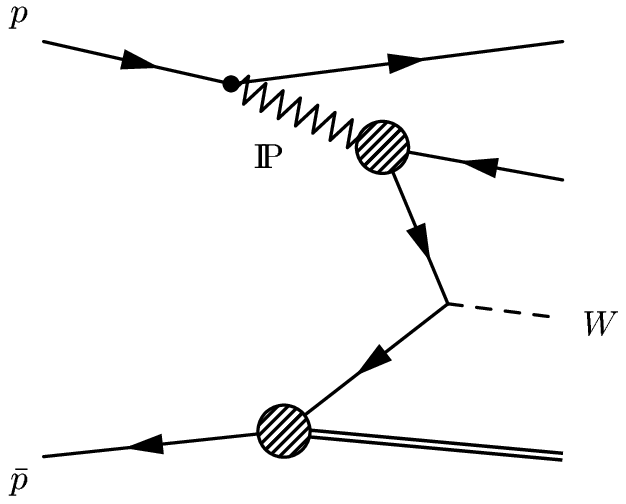,width=55mm}
\epsfig{file=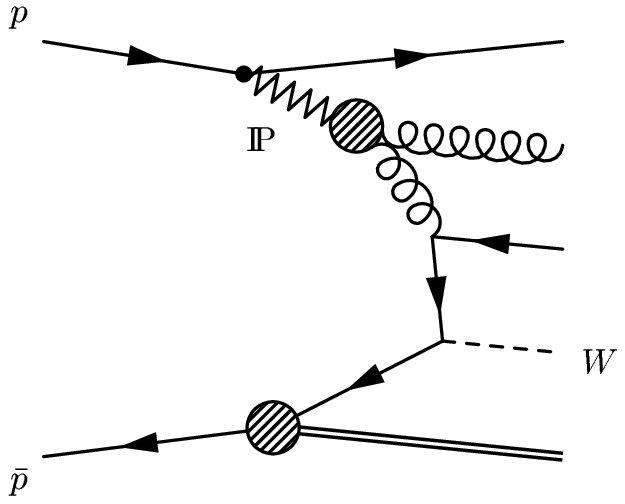,width=55mm}
\end{center}
\vspace*{-7mm}
\caption[junk]{Diffractive $W$ (or $Z$) production in $p\bar{p}$ for a 
pomeron composed of (a) $q\bar{q}$ and (b) gluons. 
\label{fig:diffW}}
\end{figure}

Based on the {\sc Pompyt} model, predictions
were also made \cite{WZprediction} for diffractive $W$ and $Z$ production 
at the Tevatron $p\bar{p}$ collider, which provides sufficient energy in 
the pomeron-proton subsystem. With partons in the pomeron this occurs through
the subprocesses $q\bar{q}\to W$ and $gq\to qW$ as illustrated in 
Fig.~\ref{fig:diffW}. The latter requires an extra QCD vertex 
$g\to q\bar{q}$ and is therefore suppressed by a factor $\alpha_s$. Thus, 
a gluon-dominated pomeron leads to a smaller diffractive $W$ cross section 
than a $q\bar{q}$-dominated pomeron. However, in both cases the cross sections
were found to be large enough to be observable and the decay products of the 
$W$ ($Z$) often emerge in a central region covered by the detectors.  
Moreover, a measurement of these decay products, ideally muons from $Z$ decay,
allows a reconstruction of the $x$-shape of the partons in the pomeron 
\cite{WZprediction}. 

Diffractive $W$ production at the Tevatron was recently observed by CDF
resulting in a diffractive to non-diffractive $W$ production ratio 
$R_W=(1.15\pm0.55)\% $ \cite{CDF-W}. Since leading protons could
not be detected, diffraction was defined in terms of a large forward 
rapidity gap, which in terms of a pomeron model corresponds 
to $x_{\Pma}$ dominantly in the range 0.01--0.05. The observed $R_W$ is 
much smaller than predicted with a $q\bar{q}$ dominated pomeron. Using 
{\sc Pompyt} with the standard pomeron flux of eq.~(\ref{eq:pomeron-flux}) 
and pomeron parton densities obtained from fits to the HERA diffractive 
DIS data, results in $R_W=5-6\% $, \ie \ several standard deviations 
above the measured value! 

Diffractive hard scattering has also been observed at the Tevatron 
in terms of rapidity gap events with two high-$p_{\perp}$ jets (dijets)
as in UA8.  
The detailed definitions of gaps and jets differ somewhat between CDF and 
D0, but the results are similar. The ratio of diffractive to non-diffractive
dijet events found at $\sqrt{s}=1800\, GeV$ 
by CDF is $R_{jj}=(0.75\pm 0.05\pm 0.09)\%$ \cite{CDF-dijets} 
and by D0 $R_{jj}=(0.76\pm 0.04\pm 0.07)\%$ \cite{D0-dijets}. 
D0 has also obtained the ratio $R_{jj}=(1.11\pm 0.11\pm 0.20)\%$
at the lower cms energy $\sqrt{s}=630\, GeV$. 
These rates are significantly lower than those obtained with the standard
pomeron model with parton densities that fit the diffractive HERA data.

\begin{figure}[hbt]
\begin{center}
\epsfig{file=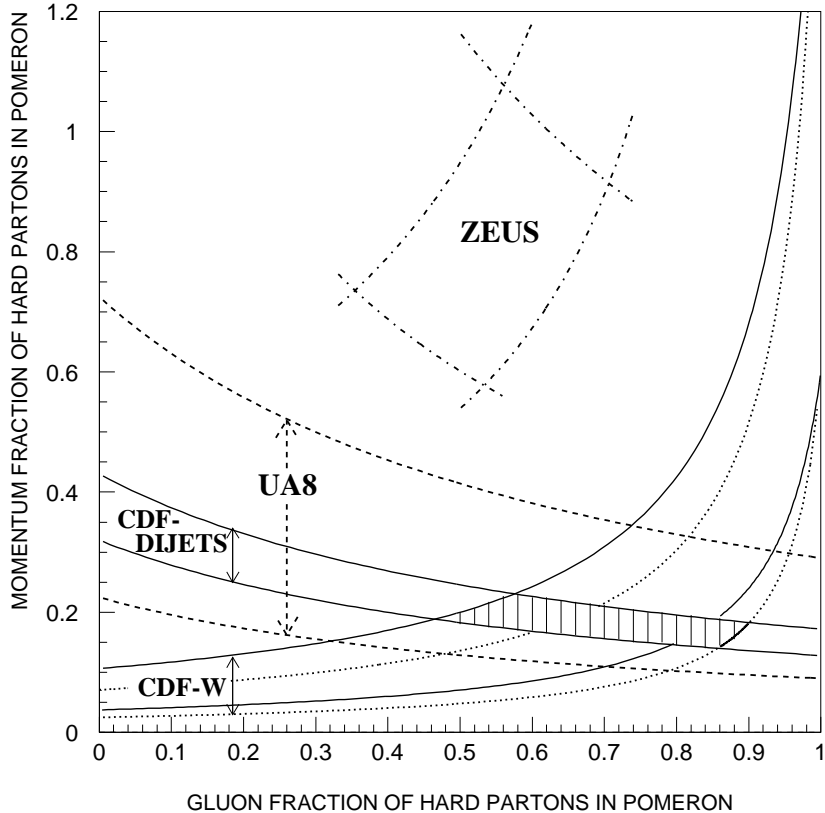,width=110mm,
   bbllx=75pt,bblly=240pt,bburx=510pt,bbury=670pt,clip=}
\end{center}
\vspace*{-7mm}
\caption[junk]{Total momentum fraction versus gluon fraction of hard partons 
in the pomeron evaluated by comparing measured diffractive rates with 
Monte Carlo results based on the standard pomeron flux showing a mismatch 
between  results from $p\bar{p}$ and $ep$.  
From \cite{CDF-dijets}.
\label{fig:pomerontest}}
\end{figure}
The inability to describe the data on hard diffraction from both HERA and 
the Tevatron with the same pomeron model raises questions on the universality 
of the model, \eg \ concerning the pomeron flux and structure function. 
This is examined in Fig.~\ref{fig:pomerontest} in terms of the momentum sum 
of the partons and the amount of gluons needed to fit the data. 
The region acceptable to HERA data is compatible with a saturated 
momentum sum rule, but in disagreement with the internally consistent
$p\bar{p}$ collider data.

CDF has also very recently observed events with a central dijet system and 
rapidity gaps on both sides. On one side a high-$x_F$ antiproton is actually 
detected. This can be interpreted as double pomeron exchange 
(\cf \ Fig.~\ref{fig:rapgaps}d),
one from each of the quasi-elastically scattered proton and antiproton, 
where the two pomerons interact to produce the jets. The diffractive hard 
scattering model then contains a convolution of two pomeron flux factors 
and two pomeron parton densities with a QCD parton level cross section. 
The observed ratio of two-gap jet events to the single-gap jet events is found 
by CDF to be $(0.26\pm 0.05 \pm 0.05)\%$ \cite{doublegaps}. An important 
observation is also that the $E_{\perp}$-spectrum of the jets in these two-gap
events have the same {\em shape} as in single-gap and no-gap events. 
This hints at the same underlying hard scattering dynamics which does not
change with the soft processes that cause gaps or no-gaps. It is not yet 
clear whether this feature appears naturally in the pomeron model. 
However, the double pomeron exchange model, with pomeron flux and parton 
densities based on diffractive HERA data, seems to overestimate 
the rate of two-gap jet events \cite{doublegaps}.

%%%%%%%%%%%%%%%%%%%%%%%%%%%%%%%%%%%%%%%%%%%%%%%%%%%%%%%%%%%%%%%%%%%%%%%%%%%%%
%
\section{Pomeron problems}\label{sec:problems}
The inability to describe both HERA and $p\bar{p}$ collider data on 
hard diffraction is a problem for the pomeron model. It shows that 
the `standard' pomeron flux factor and pomeron parton densities
cannot be used universally. A possible cure to this problem has been
proposed in terms of a pomeron flux `renormalization' \cite{fluxrenorm}. 
The flux in eq.~(\ref{eq:pomeron-flux}) is found to give a much larger 
cross section for inclusive single diffraction than measured at $p\bar{p}$ 
colliders, although it works well for lower energy data. 
This is due to the increase of $f_{\Pma}\sim 1/x_{\Pma}^{2\alpha_{\Pma}(t)-1}$ 
as the minimum $x_{\Pma\, min}=M^2_{X\, min}/s$ gets smaller with 
increasing energy $\sqrt{s}$.
To prevent that the integral of the pomeron flux increases without bound,
it is proposed that it should saturate at unity, \ie \ one renormalizes 
the pomeron flux 
by dividing with its integral whenever the integral is larger than unity. 
This prescription not only gives the correct inclusive single diffractive
cross section at collider energies, but it also makes the HERA and Tevatron 
data on hard diffraction compatible with the pomeron hard scattering model. 
The model result for HERA is not affected, but at the higher energy of the 
Tevatron the pomeron flux is reduced such that the data are essentially 
reproduced. 
In another proposal \cite{Erhan-Schlein} based on an analysis of single 
diffraction cross sections, the pomeron flux is reduced at small $x_{\Pma}$ 
through a $x_{\Pma}$- and $t$-depending damping factor. Neither of these 
two modified pomeron flux factors have a clear theoretical basis. 

A difference between diffraction in $ep$ and $p\bar{p}$ is the possibility 
for coherent pomeron interactions in the latter \cite{CollinsFrankfurtStrikman}. 
In the incoherent interaction only one parton from the pomeron participates
and any others are spectators. However, in the pomeron-proton interaction 
with $\Pma =gg$ both gluons may take part in the hard interaction
giving a coherent interaction. For example, in the $\Pma p$ hard scattering 
subprocess $gg\to q\bar{q}$, the second gluon from the pomeron 
may couple to the gluon from the proton. 
Such diagrams cancel when summing over all final states for the inclusive 
hard scattering cross section (the factorization theorem). 
For gap events, however, the sum is not over all final states and 
the cancellation fails leading to factorization breaking and 
these coherent interactions where 
the whole pomeron momentum goes into the hard scattering system. 
With momentum fraction $x$ of the first gluon and $1-x$ of the second, 
a factor $1/(1-x)$ arises from the propagator of the second, soft gluon in 
the pomeron. This may motivate a super-hard component in the pomeron
with effective structure function $1/(1-x)\approx \delta (1-x)$ as in 
the UA8 data discussed above. 
This coherent interaction cannot occur in the same way in DIS since 
the pomeron interacts with a particle without coloured constituents. 
This difference between $ep$ and $p\bar{p}$ means that there should be
no complete universality of parton densities in the pomeron. 

Although modified pomeron models may describe the rapidity gap 
events reasonably well, there is no satisfactory understanding of the 
pomeron and its interaction mechanisms. On the contrary, there are 
conceptual and theoretical problems with this framework. The pomeron is 
not a real state, but can only be a virtual exchanged spacelike object. 
The concept of a
structure function is then not well defined and, in particular, it is unclear
whether a momentum sum rule should apply. In fact, the factorisation into 
a pomeron flux and a pomeron structure function cannot be uniquely defined 
since only the product is an observable quantity \cite{Landshoff_Paris}. 

It may be incorrect to consider the pomeron as being `emitted' by the proton,  
having QCD evolution as a separate entity and being `decoupled' from the 
proton during and after the hard scattering.
Since the pomeron-proton interaction is soft, its time scale is long 
compared to the short space-time scale of the hard interaction. 
It may therefore be natural to expect soft interactions between the
pomeron system and the proton both before and after the snapshot of the 
high-$Q^2$ probe (as illustrated in Fig.~\ref{fig:P-soft}c). 
The pomeron can then not be considered as decoupled from the 
proton and, in particular, is not a separate 
part of the QCD evolution in the proton. 

Large efforts have been made to understand the pomeron as two-gluon system
or gluon ladder in pQCD. By going to the soft limit one may then hope to gain 
understanding of non-pQCD. Perhaps one could establish a connection 
between pQCD in the small-$x$ limit and Regge phenomenology. More explicitly,
attempts have been made to connect the Regge pomeron with gluon ladders 
in pQCD. For example, the analogy between the Regge triple pomeron diagram
for single diffractive scattering has been connected with the gluon ladder fan 
diagram in pQCD to estimate the pomeron gluon density \cite{Bartels+GI}.
The fan diagrams are described by the GLR equation \cite{GLR} which 
gives a novel QCD evolution with non-linear effects due to gluon recombination
$gg\to g$. This reduces the gluon density at small-$x$ (screening); 
an effect that could be substantial in the pomeron \cite{Ingelman-Prytz}. 

Diffractive DIS has been considered in terms of models based on two-gluon 
exchange in pQCD, see \eg \ \cite{pQCD-pomeron}. 
The basic idea is to take two gluons in a colour singlet state from the 
proton and couple them to the $q\bar{q}$ system from the virtual photon. 
With higher orders included the diagrams and calculations become quite 
involved. Nevertheless, these formalisms can be made to describe the main  
features of the diffractive DIS data. Although this illustrates the 
possibilities of the pQCD approach to the pomeron, one is still forced to 
include non-perturbative modelling to connect the two gluons in a soft 
vertex to the proton. Thus, even if one can gain understanding by working
as far as possible in pQCD, one cannot escape the fundamental problem 
of understanding non-pQCD. 

%%%%%%%%%%%%%%%%%%%%%%%%%%%%%%%%%%%%%%%%%%%%%%%%%%%%%%%%%%%%%%%%%%%%%%%%%%%%%
% 
\section{Non-perturbative QCD and soft colour interactions}\label{sec:SCI}
The main problem in understanding diffractive interactions is related to our 
poor theoretical knowledge about non-pQCD. The Regge approach with 
a pomeron can apparently be made to work phenomenologically, 
but has problems as discussed above. 
Therefore, new models have recently been constructed without using the 
pomeron concept or Regge phenomenology. Instead, they are based on new ideas 
on soft colour interactions that give colour rearrangements which affect the 
hadronization and thereby the final state. These models have first been 
developed for diffractive DIS which is a simpler and cleaner process than 
diffraction in $p\bar{p}$ collisions.  

One model \cite{Buchmuller-Hebecker} to understand diffractive DIS at HERA 
exploits the dominance of the photon-gluon
fusion process $\gamma^{\star}g\to q\bar{q}$ at small-$x$. The $q\bar{q}$ pair 
is produced in a colour octet state, but it is here assumed that soft 
interactions 
with the proton colour field randomizes the colour. The $q\bar{q}$ pair would 
then be in an octet or singlet state with probability 8/9 and 1/9, respectively.
When in a singlet state, the $q\bar{q}$ pair hadronizes independently of the 
proton remnant, which should result in a lack of particles in between. 
From the photon-gluon fusion matrix element one then obtains the diffractive
structure function 
% \begin{equation}\label{eq:BH} 
% F_2^D(x,Q^2,\xi)\simeq \frac{1}{9} \cdot 
% \frac{\alpha_s}{2\pi} \sum_q e_q^2 g(\xi) \cdot   
% \beta \left\{ \left[ \beta^2+(1-\beta)^2\right] \ln{\frac{Q^2}{m_g^2\beta^2}} 
% -2 + 6\beta (1-\beta)\right\} 
% \end{equation}
\begin{equation}\label{eq:BH} 
F_2^D(x,Q^2,\xi)\simeq \frac{1}{9} \cdot 
\frac{\alpha_s}{2\pi} \sum_q e_q^2 g(\xi) \cdot   
\beta \{ [ \beta^2+(1-\beta)^2 ] \ln{\frac{Q^2}{m_g^2\beta^2}} 
-2 + 6\beta (1-\beta) \} 
\end{equation}
where 1/9 is the colour singlet probability.
The next factor, including the density 
$g(\xi)$ of gluons with momentum fraction $\xi$, corresponds to a pomeron 
flux factor. The $\beta$-dependent factor corresponds to 
the pomeron structure function $F_2^D(\beta,Q^2)$ above,   
with $\beta = x/\xi$ as usual. 
Thus, there is an effective factorisation which is similar to pomeron models. 
The gluon mass parameter $m_g$ regulates the divergence in the QCD matrix 
element and is chosen so as to saturate the DIS cross section at small-$x$ with 
the photon-gluon fusion process. The model reproduces main features of the gap 
events, such as their overall rate and $Q^2$ dependence. However, it is simple 
and does not take into account higher order parton emissions and hadronization. 
Therefore, it cannot give as detailed predictions as the Monte Carlo models 
above. 

\begin{figure}[hbt]
\begin{center}
\epsfig{file=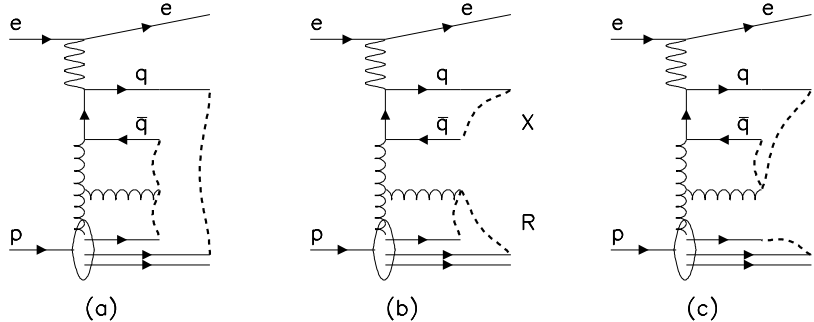,width=110mm} 
\end{center}
\vspace*{-8mm}
\caption[junk]{Gluon-induced DIS event with examples of colour string 
connection (dashed lines) of partons in 
(a) conventional Lund model based on the colour order in pQCD, and 
(b,c) after soft colour interactions.
\label{fig:SCI}}
\end{figure}

In the same general spirit another model was developed independently 
using a  Monte Carlo event generator approach \cite{SCI,Unified}. 
The starting point is the normal DIS parton interactions, with pQCD corrections
in terms of matrix elements and parton showers in the initial and final state. 
The basic new idea is that there may be additional soft colour 
interactions (SCI) between the partons at a scale below the cut-off $Q_0^2$ for the 
perturbative treatment. Obviously, interactions will not disappear below this
cut-off, the question is rather how to describe them properly.  
The proposed SCI mechanism can be viewed as the perturbatively produced 
quarks and gluons interacting softly with the colour medium of the proton
as they propagate through it. This should be a natural part of the process 
in which `bare' perturbative partons are `dressed' into non-perturbative ones
and the formation of the confining colour flux tube in between them. These 
soft interactions cannot change the momenta of the partons significantly, but 
may change their colour and thereby affect the colour structure of the 
event. This corresponds to a modified topology of the string in the Lund model 
approach, as illustrated in Fig.~\ref{fig:SCI}, such that another final state 
will arise after hadronization. 

Lacking a proper understanding of non-perturbative QCD processes a simple model 
was constructed to describe and simulate soft colour interactions. 
The hard parton level interactions are treated in the normal way using the 
{\sc Lepto} Monte Carlo \cite{Lepto} based on the standard electroweak 
cross section together with pQCD matrix elements and parton showers.
The perturbative parts of the model are kept 
unchanged, since these hard processes cannot be altered by softer 
non-pQCD ones.
Thus, the set of partons, including the quarks in the proton remnant, 
are generated as in conventional DIS. 
The SCI model is added by giving each pair of 
these colour charged partons the possibility to make a soft interaction, 
changing only the colour and not the momentum. This may be viewed as soft 
non-perturbative gluon exchange. 
Being a non-perturbative  process, the exchange probability cannot be 
calculated and is therefore described by a phenomenological parameter $R$.    
The number of soft exchanges will vary event-by-event and change the colour
topology such that, in some cases, colour singlet subsystems arise separated
in rapidity as shown in Fig.~\ref{fig:SCI}bc. Here, (b) can be seen as a switch 
of anticolour between the antiquark and the diquark and (c) as a switch of 
colour between the two quarks. Colour exchange between the perturbatively 
produced partons and the partons in the proton remnant
(representing the colour field of the proton)  
are of particular importance for the gap formation.

Both gap and no-gap events arise in this model. 
The rate and main properties of the gap events 
are qualitatively reproduced \cite{Unified}, 
\eg \ the $\eta_{max}$ distribution in Fig.~\ref{fig:gap_ZEUS_etamax}b
and the diffractive structure function $F_2^{D(3)}$. 
The gap rate depends on the parameter $R$, but the dependence is not strong 
giving a stable model with $R\simeq 0.2$--0.5. 
This colour exchange probability is the only new parameter in the model. 
Other parameters belong to
the conventional DIS model \cite{Lepto} and have their usual values. 
The rate and size of gaps do, however, depend on the amount of parton 
emission. In particular, more initial state parton shower emissions will tend 
to populate the forward rapidity region and prevent gap formation
\cite{Unified}.  

\begin{figure}[hbt]
\begin{center}
\epsfig{file=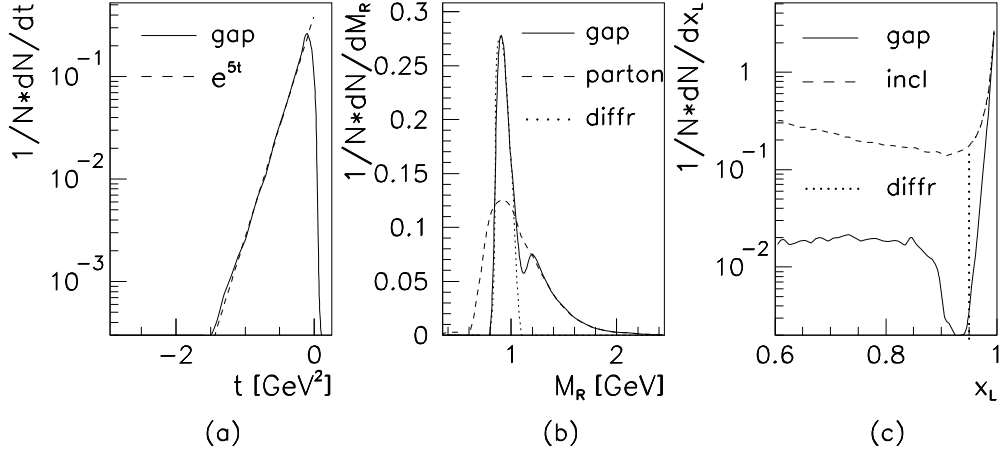,width=140mm} 
\end{center}
\vspace*{-8mm}
\caption[junk]{(a) Squared momentum transfer $t$ from initial proton to remnant
system $R$ for `gap' events compared with the exponential slope 
$1/\sigma_i^2 =5\: GeV^2$. (b) Invariant mass $M_R$ of the forward remnant 
system for `gap' events at the hadron level (solid line) and parton level 
(dashed line) compared with `diffractive' events at the hadron level (dotted
line). (c) Longitudinal momentum fraction $x_L$ for final protons in 
`gap' events (solid line), all events (dashed line) and 
`diffractive' events (dotted line). 
From \cite{Unified}
\label{fig:remnant}}
\end{figure}

The gap events show properties characteristic of 
diffraction as demonstrated in Fig.~\ref{fig:remnant}. 
The exponential $t$-dependence arises in the model 
from the gaussian intrinsic transverse momentum (Fermi motion) of the 
interacting parton which is balanced by the proton remnant system, 
\ie \ $exp(-k_{\perp}^2/\sigma_i^2)$ with $\sigma_i\simeq 0.4\: GeV$ and 
$t\simeq -k_{\perp}^2$. 
The forward system (Fig.~\ref{fig:remnant}b) is dominantly a single proton, 
as in diffractive scattering, but there is also a tail corresponding to 
proton dissociation. 
The longitudinal momentum spectrum of protons in Fig.~\ref{fig:remnant}c 
shows a clear peak at large fractional momentum $x_L$. Defining events having
a leading proton with $x_L > 0.95$ as `diffractive', one observes in 
Fig.~\ref{fig:remnant}bc that most of these events fulfill the gap 
requirement.   

One may ask whether this kind of soft colour interaction model is essentially
a model for the pomeron. This is not the case as long as no pomeron or 
Regge dynamics is introduced. 
The behaviour of the data on $F_2^D(\beta,Q^2)$ in Fig.~\ref{fig:F2Dtildeqg}
is in the SCI model understood as normal pQCD evolution in the proton. 
The rise with $ln Q^2$ also at larger $\beta$ is simply the normal behaviour
at the small momentum fraction $x=\beta x_{\Pma}$ of the parton in the proton.
Here, $x_{\Pma}$ is only an extra variable related to the gap size or 
$M_X$ (eq.~(\ref{eq:invariantdef})) which does not require a pomeron 
interpretation. The flat $\beta$-dependence (Fig.~\ref{fig:F2Dtildeqg}b)
of $x_{\Pma} F_2^D=\frac{x}{\beta} F_2^D$ is due to the factor $x$
compensating the well-known increase at small-$x$ of the proton structure 
function $F_2$. 
 
This Monte Carlo model gives a general description of DIS, with and without 
gaps. In fact, it can give a fair account for such `orthogonal' observables as 
rapidity gaps and the large forward $E_{\perp}$ flow \cite{Unified}. 
Diffractive events are in this model defined through the topology of the final 
state, in terms of rapidity gaps or leading protons just as in experiments.  
There is no particular theoretical mechanism or description in a separate 
model, like pomeron exchange, that defines what is labelled as diffraction. 
This provides a smooth transition between diffractive gap events and 
non-diffractive no-gap events \cite{SCI-forward}. 
In addition, leading neutrons are also obtained in fair agreement with 
recent experimental measurements \cite{leading-pn}. In a conventional 
Regge-based approach, pomeron exchange would be used to get diffraction, 
pion exchange added to get leading neutrons and still other exchanges added 
to get a smooth transition to normal DIS. The SCI model indicates that a 
simpler theoretical description can be obtained.  

\begin{figure}[hbtp]
\begin{center}
\epsfig{file=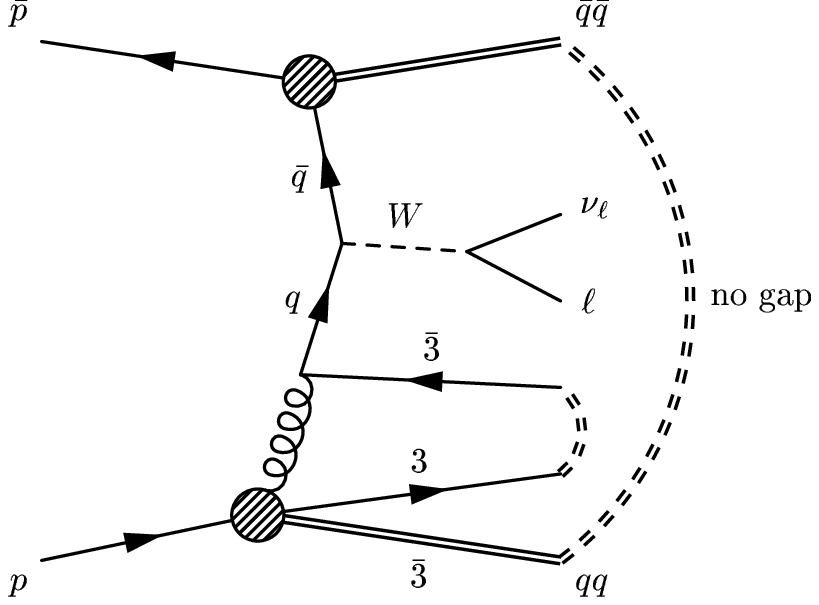,width=65mm}
\epsfig{file=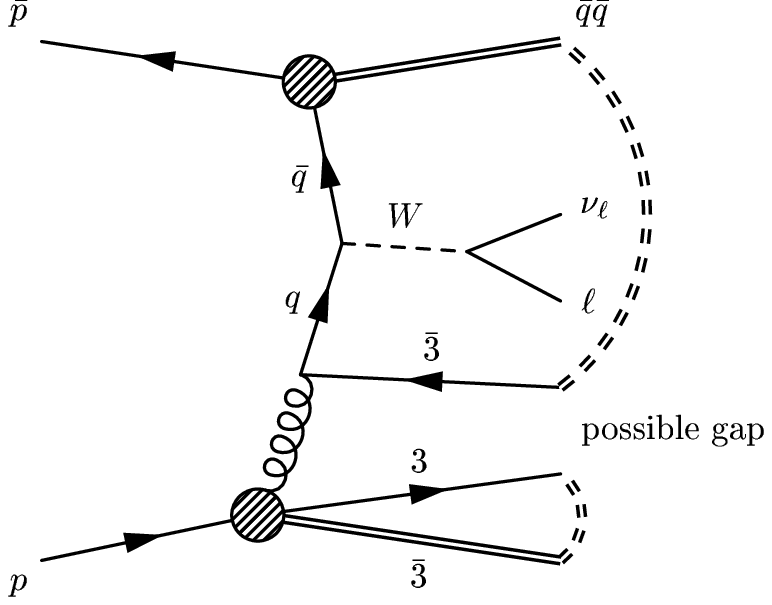,width=65mm}
\end{center}
\vspace*{-7mm}
\caption[junk]{$W$ production in $p\bar{p}$ with examples of colour string 
connections (dashed lines) of partons in (a) the conventional {\sc Pythia}
model and (b) after soft colour interactions. 
\label{fig:SCI-W}}
\end{figure}

The same SCI model can also be applied to $p\bar{p}$ collisions, 
by introducing it in the {\sc Pythia} Monte Carlo \cite{Pythia}. 
This leads to gap events in hard scattering interactions as
illustrated for $W$ production in Fig.~\ref{fig:SCI-W}.
It is amazing that the same SCI model, normalized to the diffractive 
HERA data, reproduces the above discussed rates of diffractive $W$'s and 
diffractive jet production observed at the Tevatron \cite{SCI-W-jets}. 

The soft colour interactions do not only lead to rapidity gaps, but also 
to other striking effects. They have been found \cite{SCI-psi} to reproduce 
the observed rate of high-$p_{\perp}$ charmonium and bottomonium at the 
Tevatron, which are factors of 10 larger than predictions based on 
conventional pQCD. The SCI model included in {\sc Pythia} 
accomplish this through the standard pQCD parton level processes of 
heavy quark pair production.
The most important contribution comes from a high-$p_\perp$ gluon
which splits in a $Q\bar{Q}$ pair, \eg \ the next-to-leading process 
$gg \to gQ\bar{Q}$, where the colour octet charge of the $Q\bar{Q}$ 
can be turned into a singlet through SCI. The $Q\bar{Q}$ pairs with 
mass below the threshold for open heavy flavour production, are then mapped 
onto the various quarkonium states using spin statistics. 
The results \cite{SCI-psi} are in good agreement with the data, 
both in terms of absolute normalization and the shapes. 
Also details like the rates of different quarkonium states and the fraction 
of $J/\psi$ produced directly or from decays are reproduced quite well.  

This simple model for soft colour interactions is quite 
successful to describe a lot of data, both for diffractive and 
non-diffractive events. Of course, it is only a very simple model 
and far from a theory, but it may lead to a proper description. 
A very recent step in this direction is the use of an area law 
for string dynamics \cite{Rathsman}. 

The SCI model has similarities with other attempts to 
understand soft dynamics. 
Soft interactions of a colour charge moving through a colour medium has 
been considered and argued to give rise to large $K$-factors in Drell-Yan 
processes and synchrotron radiation of soft photons \cite{ColourMedium}. 
A semi-classical approach to describe the interaction of a $q\bar{q}$ pair 
with a background colour field of a proton has been developed into a model 
for diffraction in DIS \cite{Buchmuller-Hebecker2}. The $q\bar{q}$, which 
is here a fluctuation of the exchanged virtual photon, can emerge in 
a colour singlet state after the interaction with the proton such that a
rapidity gap can arise. This provides a very interesting theoretical 
framework giving results in basic agreement with data, although one cannot 
make as detailed comparisons as with a Monte Carlo model. 

Other attempts to gain understanding through phenomenological models have 
also been made in the same general spirit as the SCI model.
The colour evaporation model \cite{Halzen} can reproduce rapidity gap data 
and charmonium production with fitted parameters to regulate the probability
of forming colour singlet systems. 
Changes of colour string topologies have also been investigated 
\cite{StringReconn} in a different context, namely 
$e^+e^- \rightarrow W^+W^- \rightarrow q_1\bar{q}_2q_3\bar{q}_4$. 
This gives two strings that may interact and cause colour reconnections 
resulting in a different string topology affecting the $W$ mass reconstruction 
and Bose-Einstein effects. 

In conclusion, there has been an increased interest in recent years 
to explore non-pQCD through various theoretical attempts and phenomenological
soft interaction models. 

%%%%%%%%%%%%%%%%%%%%%%%%%%%%%%%%%%%%%%%%%%%%%%%%%%%%%%%%%%%%%%%%%%%%%%%%%%%%%
%
\section{Rapidity gaps between jets}\label{sec:jgj}
The diffractive events discussed so far always had a rapidity gap adjacent to 
a leading proton or small mass system. The momentum transfer between the 
initial proton and this very forward system is always very small (exponential
$t$-distribution) as characteristic of soft processes. This applies whether
the high-mass $X$-system contains hard scattering or not. 
In $p\bar{p}$ collisions at the Tevatron one has discovered a new kind of
rapidity gaps, namely where the gap is in the central region and between 
two jets with high $p_\perp$, \ie \ `jet-gap-jet' events. 

In a sample of $p\bar{p}$ events at $\sqrt{s}=1800\, GeV$ having two jets 
with transverse energy $E_\perp^{jet} >20 \, GeV$, pseudorapidity 
$1.8<|\eta_{jet}|<3.5$ and $\eta_{jet1}\eta_{jet2}<0$, 
CDF finds \cite{CDF-jgj} that a fraction $R_{jgj}=(1.13\pm 0.12 \pm 0.11)\%$
has a rapidity gap within $|\eta |<1$ between the jets. 
At $\sqrt{s}=630\, GeV$ the CDF result is $R_{jgj}=(2.7\pm 0.7 \pm 0.6)\%$
with $E_\perp^{jet} >8\, GeV$ which corresponds to approximately the same 
momentum fraction $x$ of the interacting partons at the two cms energies.
D0 finds \cite{D0-jgj} 
very similar results in terms of `colour singlet fractions'
$f_s=(0.94\pm 0.04 \pm 0.12)\%$ for $E_\perp^{jet} >30 \, GeV$ at 
$\sqrt{s}=1800\, GeV$ and 
$f_s=(1.85\pm 0.09 \pm 0.37)\%$ for $E_\perp^{jet} >12 \, GeV$ at 
$\sqrt{s}=630\, GeV$. 
Although the CDF and D0 event selections and analyses differ, 
the resulting relative rates of jet-gap-jet events are quite similar. 
They are definitely larger at the lower energy. 
In D0 the ratios tend to increase with increasing $E_\perp^{jet}$ and rapidity 
separation between the jets, but the CDF data shows no significant such effect.

The jet-gap-jet events can be interpreted in terms of colour singlet 
exchange. However, the momentum transfer 
$|t|\sim E^2_{\perp jet}> 100\, GeV^2$ is very large in contrast 
to the small $t$ in ordinary diffraction. An interpretation in terms
of the Regge pomeron is therefore not possible, but attempts have been 
made using pQCD models of two-gluon exchange. Such models seems at first
to give energy and $E_\perp^{jet}$ dependences that are not consistent with 
the data, but recent developments indicate that this need not be the case
\cite{Zeppenfeld}. 
The salient features of the data can, on the other hand, be interpreted 
in terms of the colour evaporation model \cite{Halzen-jgj}. A problem with 
both these approaches is however, that they do not take proper account of 
higher order pQCD parton emissions, multiple parton-parton scattering and 
hadronization. These are well known problems for the understanding 
of the `underlying event' in hadron-hadron collisions and must be investigated 
with detailed Monte Carlo models. For example, the perturbative radiation in 
a high-$p_\perp$ scattering must be included since it cannot be screened 
by soft interactions. The proposed models attempts to describe all these 
effects through a `gap survival probability' \cite{Bj}. 
However, a real understanding of gap between jets is still lacking. 

%%%%%%%%%%%%%%%%%%%%%%%%%%%%%%%%%%%%%%%%%%%%%%%%%%%%%%%%%%%%%%%%%%%%%%%%%%%%%
%
\section{Conclusions}\label{sec:conclusions}
Diffractive hard scattering has in recent years been established as a 
field of its own with many developments in both theory and experiment.
Rapidity gap events have been observed with various hard scattering 
processes;
high-$p_\perp$ jet and $W$ production, and deep inelastic scattering. 

The model with a pomeron having a parton structure is quite successful 
in describing data, in particular for diffractive DIS at HERA where parton 
densities in the pomeron have been extracted. However, the pomeron model
has some problems. The pomeron flux and/or the pomeron parton densities 
are not universal to all kinds of interactions, or they are more 
complicated with, \eg , a flux renormalization. Even if such modified 
pomeron models can be made to describe data both from $ep$ and $p\bar{p}$, 
there are conceptual problems with the pomeron. 
In particular, it is doubtful whether the pomeron can 
be viewed as a separate entity which is decoupled from the proton during 
the long space-time scale of the soft interaction. 

The general problem is soft interactions in non-perturbative 
QCD. Perhaps Regge theory is the proper soft limit of QCD, but it may also
exist more fruitful roads towards a theory for soft interactions.
This has generated an increased interest to 
explore new theoretical approaches and phenomenological models.  

A new trend is to consider the interactions of partons with a 
colour background field. The hard pQCD processes should then be 
treated as usual, but soft interactions are added which change the colour 
topology resulting in a different final state after hadronization. 
In the Monte Carlo model for soft colour interactions this gives a 
unified description with a smooth
transition between diffractive and non-diffractive events. The different 
event classes can then be defined as in experiments, \eg \ in terms of 
rapidity gaps or leading protons. 
This model and others in a similar general spirit can describe 
the salient features of many different kinds of experimental data. 

Nevertheless, there are many unsolved problems that are challenging to solve.
In particular, the events with a rapidity gap between two high-$p_\perp$ jets
are poorly understood. Progress in the field of diffractive hard scattering 
will contribute to the ultimate goal: to understand non-perturbative QCD. 

\vspace*{2mm}
\noindent
{\bf Acknowledgments:}
I am grateful to Tom Ferbel and all the participants for a most 
enjoyable school. 

\end{document}